\documentclass[9pt,twocolumn,twoside]{IEEEtran}

\usepackage{subcaption}

\usepackage{placeins}
\usepackage{graphicx}
\usepackage{hyperref}

\begin{document}

\title{The Psychophysics of Human Three-Dimensional Active Visuospatial Problem-Solving}

\author{Markus D. Solbach \and John K. Tsotsos}

\author{
\IEEEauthorblockN{Markus D. Solbach, John K. Tsotsos\\
Department of Electrical Engineering and Computer Science \\ 
York University\\
\emph{\{solbach, tsotsos\}@yorku.ca}}
}

\maketitle

\begin{abstract}

Our understanding of how visual systems detect, analyze and interpret visual stimuli has advanced greatly. However, the visual systems of all animals do much more; they enable visual behaviours. How well the visual system performs while interacting with the visual environment and how vision is used in the real world have not been well studied, especially in humans. It has been suggested that comparison is the most primitive of psychophysical tasks. Thus, as a probe into these active visual behaviours, we use a same-different task: are two physical 3D objects visually the same? This task seems to be a fundamental cognitive ability. We pose this question to human subjects who are free to move about and examine two real objects in an actual 3D space. Past work has dealt solely with a 2D static version of this problem. We have collected detailed, first-of-its-kind data of humans performing a visuospatial task in hundreds of trials. Strikingly, humans are remarkably good at this task without any training, with a mean accuracy of 93.82\%. No learning effect was observed on accuracy after many trials, but some effect was seen for response time, number of fixations and extent of head movement. Subjects demonstrated a variety of complex strategies involving a range of movement and eye fixation changes, suggesting that solutions were developed dynamically and tailored to the specific task.

\end{abstract}

\section{Introduction}\label{sec:intro}

Human visual ability feels so effortless that it is literally taken for granted. However, any intuitive introspection into this ability rarely reveals its profound nature. On the contrary, the tendency has been to prefer simpler descriptions in an Occam's Razor (or parsimony) sense, and although these have helped move our understanding along, they are no longer as useful. A main focus has been on how well the eyes and visual system can detect a target stimulus, i.e. Visual Function \cite{Bennett2019}. This has been studied extensively over many decades, in several literatures (in computer vision, in ophthalmology, in visual psychophysics). The visual systems of all animals do much more than simply detect stimuli; they also enable locomotion, seek food, detect threats, guide mating and rearing of offspring. In humans, there is even more than supporting basic survival. We use our visual abilities to create, exploit, admire, destroy, and manipulate our 3D physical world. How well we perform while interacting with the visual environment and how vision is used in everyday activities has been termed Functional Vision \cite{Bennett2019}. In contrast with visual function, this has not been well examined, in part due to its inherent experimental difficulty and to the fact that it seems far more complex an activity to define. Research of this kind has recently become possible in a variety of animals \cite{Parker2022a,Kadohisa2022,Martinho2016}. We report a first detailed and precise investigation into human functional vision.

\begin{figure}[h!]
  \centering
  \includegraphics[width=1.0\linewidth]{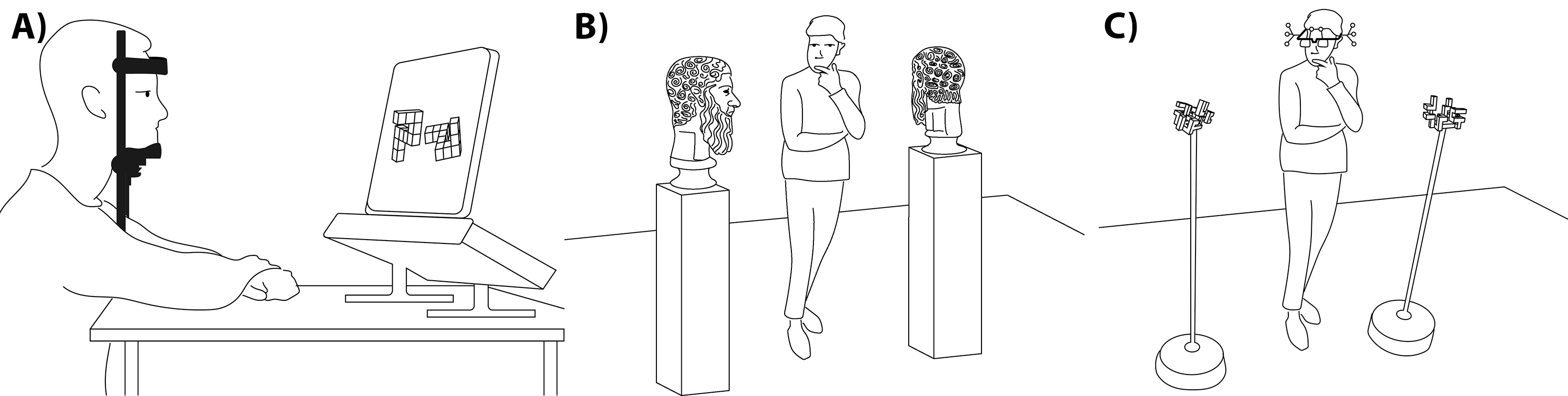}
  \caption[]{Our experiment as it evolved, beginning with Sheppard \& Metzler's inspiration, the reality of visuospatial problem-solving ``in the wild'', and our actual setup. \textit{\textbf{A)}} The study of human vision often involves a two-dimensional probe, i.e., visual function. As illustrated here, the subject sits at a desk with head stabilized in a forehead and chin mount while performing an experiment involving images on a screen. \textit{\textbf{B)}} Real-world visuospatial problem-solving, i.e. functional vision, however, is distinctively different. Here we show someone in a sculpture gallery wondering if these two marble heads are the same. Humans do not passively receive stimuli; rather, they choose what to look at and how. We move our head and body in a three-dimensional world and view objects from directions and positions that are most suited to our viewing purpose. It should be clear that answering our sculpture query here might require more than one glance. It is we who decide what to look at, not an experimenter. \textit{\textbf{C)}} Our experimental setup is as shown. A subject wears a special, wireless headset and is shown two objects mounted on posts at certain 3D orientations. The subject is asked to determine, without touching, if the two objects are the same (in all aspects) or different and is completely free and untethered to move anywhere they choose. Gaze and view are precisely recorded (See Methods \& Materials).}
  \label{fig:functional_vision}
\end{figure}

Although it seems easy to enumerate the many kinds of behaviours humans perform in their visual world, it is far less easy to know how to best probe the nature of such behaviours in a general sense. A classic example is seen in the well-known experiments of Sheppard \& Metzler in 1971 \cite{Shepard2019} where they showed subjects, seated in front of a video monitor, 2D projections of unknown 3D geometric objects (Figure \ref{fig:functional_vision}A). Subjects were instructed to determine if the two objects were the same or different. Note how self-occlusion is a natural characteristic of their objects. Their results showed that subjects seemed to mentally rotate objects in order to test potential geometric correspondence. This work was seminal in this area but seemed difficult to generalize if viewing real 3D objects in an unconstrained manner. Nevertheless, their experiment provided the inspiration for many important investigations as well as for what we present here. In a more foundational manner, the act of comparison has been suggested as the most primitive psychophysical task \cite{macmillan2004detection}; efforts to discover the deep nature of real-world visual comparison behaviour may have an impact on the full spectrum of human visual behaviour.

How do humans decide if two objects are the same or different (as Sheppard \& Metzler asked) within a physical 3D environment where they are free to explore those objects? Where do they look? How do they look? How do they move? Such questions are central to this study. Figure \ref{fig:functional_vision}B shows an example of one common such setting within a sculpture gallery: are these two sculptures the same? This may be easy, or it may be difficult to determine. What is clear is that a single image of the objects generally will not suffice. Suppose we consider 3D versions of Sheppard \& Metzler's unknown geometric objects. We could construct a range of such objects of varying physical complexity and show them to subjects. They would be free to examine without touching them and free to move around the objects, just like in the sculpture gallery. This requires a means to precisely record exactly what subjects are looking at and from what viewing position while deciding the answer to the simple question: are two objects the same or different? Figure \ref{fig:functional_vision}C gives a sketch of our experiment showing a human subject wearing the recording device.

With this question in mind, we have designed a novel set of stimuli with known geometric complexity. We also employ a first-of-its-kind experimental setup that allows for precise, synchronized tracking of head motion with 6 degrees of freedom and eye gaze while subjects are completely untethered to allow natural task execution. We have collected detailed data on humans performing a visuospatial task in hundreds of experiments and present an in-depth analysis with respect to various cumulative performance metrics such as the number of fixations, response time, amount of movement, and learning effect. We are not restricted to summary statistics of single actions such as these, but we also examine sequences of actions that reveal a fascinating array of problem-solving strategies employed differently by each subject and for each test case.

We also challenge the current intuitive views that perhaps such a visual ability can be captured by a large dataset of trials that feed a sophisticated learning method or that clever use of visual saliency maps might be the key, or that novel policies for a reinforcement learning approach might suffice. Our expectation is that this, and related non-trivial tasks, need something different. Our experimental goal is to reveal what human solutions to such 3D visuospatial tasks actually involve.

To analyze the effect of stimulus complexity, our set of objects consists of 12 objects divided into three complexity levels each of which can be presented at an arbitrary 3D pose. How each object appears (the image it projects to the subject) changes, sometimes radically, with change in viewing position. As a result we include not only the object poses as experimental variables but also the subject’s initial viewing position. To keep the experiment tractable we investigate three object pose orientations and three viewer starting positions. Every subject performed 18 trials with randomly selected experimental configurations to test if a learning effect exists.

Surprisingly, humans have virtually no difficulty with this task, even for hard cases. The accuracy ranged from 80-100\% across all configurations. Much data acquisition occurs with a minimum of 6 and a maximum of 800 eye fixations. Interestingly, no statistical change was observed in accuracy throughout the trials. However, a learning effect was seen for the number of fixations on the objects, response time, and head movement. The sequence of actions we observed strongly suggests that human problem-solving strategies are dynamically determined and deployed in a seemingly directed hypothesize-and-test manner tailored to the particular task instance at hand. Subjects do not need to learn the task; they develop good solutions from the start, and over the set of trials, those solutions become smoother or more efficient while maintaining accuracy.

\section{Materials and Methods}

\subsection{Stimuli and Task}

The task, including the stimuli, is illustrated in Figure \ref{fig:functional_vision}C and is designed to be a two-alternative forced choice. Subjects were allowed to move within a constrained area of about 3.4m by 4.3m, presented with two static three-dimensional stimuli mounted on acrylic posts. The task was to determine whether the two stimuli were the same or different. Sameness in our experiment is defined as geometric congruence -- all stimuli share the same colour and surface texture. 

\begin{figure}[th!]
\begin{center}
   \includegraphics[width=1.0\linewidth]{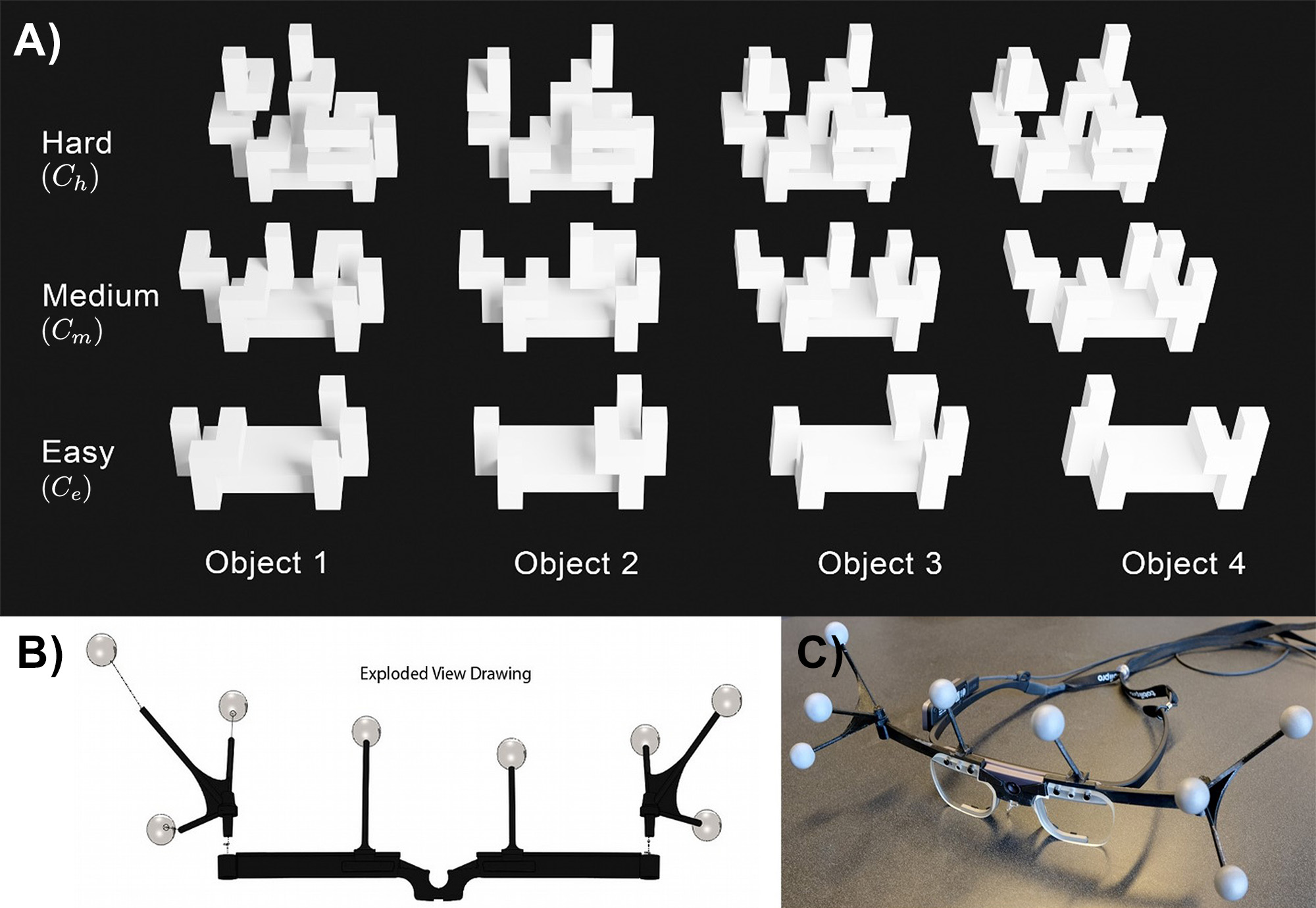}
\end{center}
   \caption{\textit{\textbf{A)}} Illustration of \textit{TEOS} objects used as stimuli. The set is split into three different complexity levels. Complexity is defined as the number of blocks used to build an object. \textit{\textbf{B)}} Expanded view drawing of the custom clip-on tracking equipment. It uses 8 rotationally variant positioned tracking markers to avoid ambiguities. \textit{\textbf{C)}} Photograph of the assembled eye tracking glasses with tracking equipment.}
\label{fig:stimData}
\end{figure}

The stimuli are part of a three-dimensional physical objects set called \textit{TEOS} \cite{Solbach2021}. The objects are inspired by the stimuli of Sheppard \& Metzler. \textit{TEOS} objects are all three-dimensional and have a known geometric complexity. Furthermore, a shared common-coordinate system allows quantifying the orientational difference of two objects. An illustration of the objects is shown in Figure \ref{fig:stimData}A. The set contains twelve objects split equally into three different complexity levels, which is defined by the number of blocks used to build an object. The level of object complexity $C$ will be indicated by the subscript, such as $C_e$, $C_m$, and $C_h$ for easy, medium, and hard, respectively. \textit{TEOS} objects are 3D printable. The objects are roughly $12cm \times 14cm \times 18cm$ in size. Movements of the subject were not restricted, and no time constraints were given. However, a definite answer (same, different, i.e. 2AFC) must be given to end the trial. Each subject performed 18 trials evenly split among complexity levels. 

\begin{figure}[th!]
\begin{center}
   \includegraphics[width=1.0\linewidth]{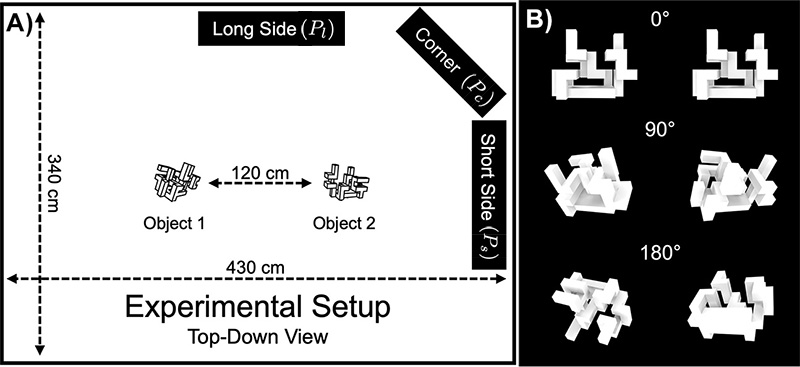}
\end{center}
   \caption{\textit{\textbf{A)}} Top-down illustration of the experimental setup. It shows dimensions, as well as the three different starting positions investigated. \textit{\textbf{B)}} We have investigated three orientational differences between the stimuli. $0^{\circ}$ (top) means that there is no rotational difference between both object poses -- the object rotations are aligned. $90^{\circ}$ and $180^{\circ}$ means that the poses have a rotational difference of $90^{\circ}$ and $180^{\circ}$, respectively. For all trials, we have used the same poses -- as shown in this illustration.}
\label{fig:tech}
\end{figure}

We also investigated different starting positions as they determine the initial observation of the objects. Figure \ref{fig:tech}A illustrates a top view of the experimental space marking the three positions from which a trial can start: equidistant from both objects ($P_l$), in line with both objects ($P_s$) and oblique to both ($P_c$).

Furthermore, we looked at the effect of object orientation difference. We limited the large space of possibilities to three values of orientation difference, $0^{\circ}$, $90^{\circ}$, and $180^{\circ}$. For all trials, we have used the same three poses as illustrated in Figure \ref{fig:tech}B. Furthermore, all experimental variables were selected randomly for each trial, including sameness, complexity, starting position, and object orientation. 

Lastly, after the experiment, subjects answered questions about their approach to solving this task (e.g. ``what was your strategy for approaching the task'', ``did you notice any changes in your approach throughout the trials'', ``which instances were more challenging than others and why'').

\subsection{Data Acquisition and Analysis}

We created a novel active vision experimental facility (named PESAO - Psychophysical Experimental Setup for Active Observers \cite{Solbach2020}) which formed the basis for all data acquisition and analysis. Its primary components are: Eye tracking glasses to capture the gaze direction, a motion tracking system for head tracking, $1^{st}$ and $3^{rd}$ person video and homogenous lighting set up. Specialized software synchronized and aligned data streams from each source at microsecond precision \cite{Solbach2020}. To track the position and orientation of the stimuli, we developed motion-tracking markers that were attached to the stand of the objects. The subject's head motion was tracked using a custom tracking body attached to the eye-tracking glasses. Figure \ref{fig:stimData}B shows the custom clip-on equipment and Figure \ref{fig:stimData}C shows a photo of the assembly with the glasses. The tracking frequency for objects and head motion was 120 Hz, and for the eyes 50 Hz. The accuracy for the motion tracking system was $\approx0.2mm$ (RMSE) in 97\% of the capture volume. The gaze was tracked with $1.42^{\circ}$ mean accuracy. Lastly, our statistical significance analysis is performed using one-way repeated-measures ANOVA.

\subsection{Subjects}

47 participants randomly sampled from the general public took part in our experiment. The average age was 23.4 years, ranging from 19 to 52 years of age. All subjects had normal, or corrected-to-normal vision, granted informed consent and were paid for participation. The experiment was approved by the office of research ethics at York University (Certificates \#2020-137 and \#2020-217).

\section{Results}

\begin{figure}[th!]
\begin{center}
   \includegraphics[width=1.0\linewidth]{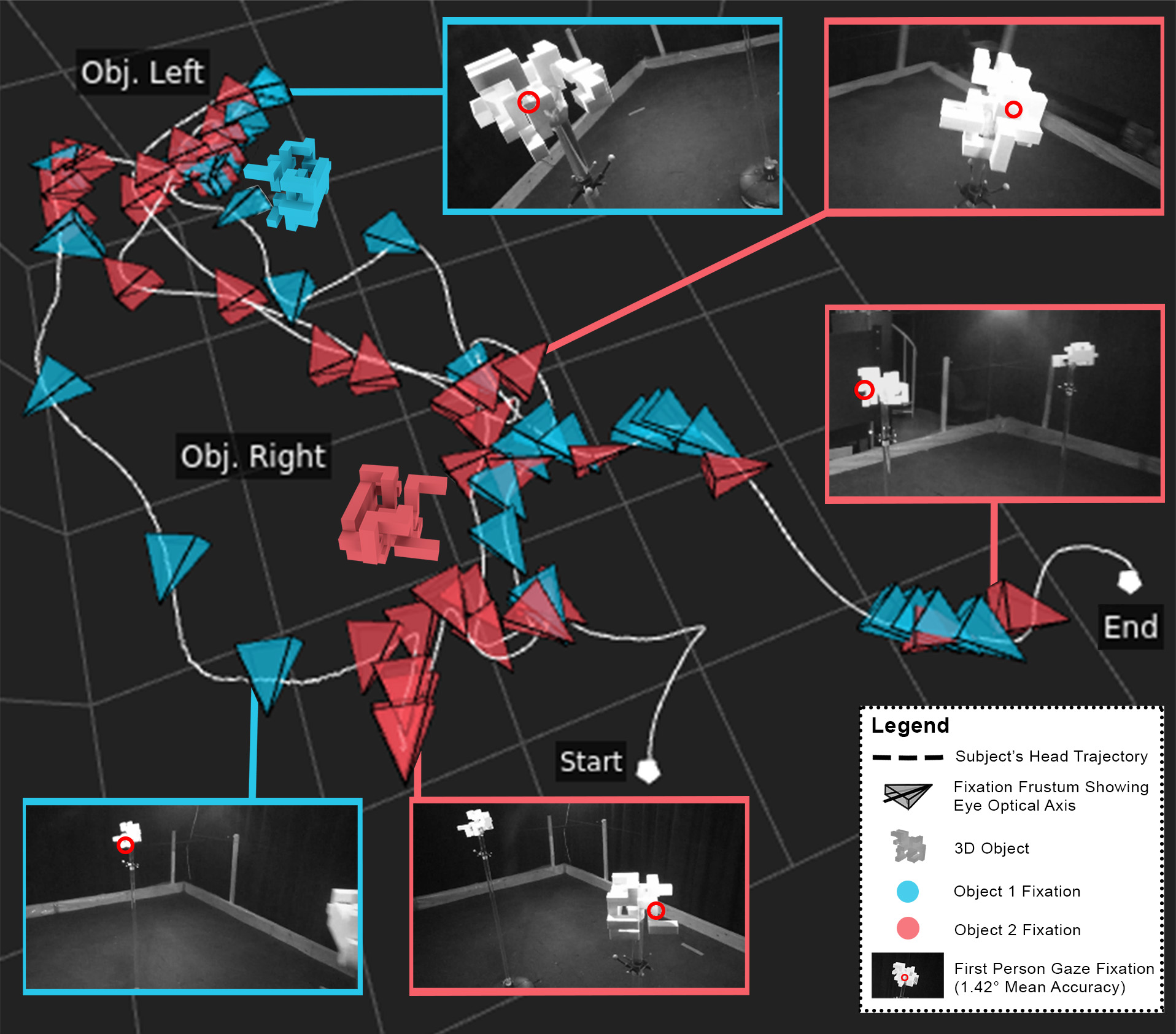}
\end{center}
   \caption{A visualization of the recorded data from PESAO. The movement of the subject is plotted as a dashed line in white, and fixations on either object are illustrated as a frustum in the corresponding colour of the fixated object. Selected fixation frusta are annotated with snapshots of the subject’s first-person view and the gaze at a particular fixation (red circle). In this example, the objects are the same, of complexity level $C_h$, they differ in pose by $180^{\circ}$, and the subject started from position $P_s$.}
\label{fig:trajectory}
\end{figure}

In total, we conducted 846 trials. Each subject performed 18 trials, sampling each configuration of experimental variables more than 15 times. We recorded about $80,000$ fixations with over 4.5M head poses and 11 hours of footage of $1^{st}$ and $3^{rd}$ person video each. A visualization of a trial is shown in Figure \ref{fig:trajectory}, visualized using the graphical functions of \textit{PESAO}. The subject required 61 fixations, moved a total of 18.75 m to complete the task, and answered correctly (same).

We next present observations on accuracy, number of fixations, response time, movement, and fixation patterns.

\subsection{Accuracy}

Humans are remarkably good at this task. Throughout all configurations, participants achieved an absolute mean accuracy of 93.83\%, $\sigma = 3.9\%$ (Figure S1, Supporting Information (SI)). The best-performing configuration was with stimuli of $C_e$, starting position $P_l$ and a difference in object orientations of $0.0^{\circ}$. Not a single trial of this configuration was answered incorrectly, regardless of the object sameness. Object complexity plays a relevant role in how well participants performed this task. Objects of $C_e$ complexity yield an average accuracy of 96.1\%, $C_m$ objects with 94.18\%, and $C_h$ with 91.2\%.

The sameness of objects has a significant effect on accuracy ($F_{1,46} = 3.58, p = 0.044$, see Figure S1 c), SI). If the objects are the same, in general, a higher accuracy is seen (94.3\%) than for different pairings (91.6\%). 

We were also interested in investigating the effect of the starting position (Figure S1 a), SI). While for the $C_e$, the best mean performance was observed from $P_c$, for $C_m$, the best performance was seen starting from $P_s$, and finally, for $C_h$, starting from $P_l$ achieved the highest accuracy. While the worst performances varied between $P_s$ and $P_c$ starting position, $P_l$ did seem to result, generally, in higher accuracy. However, no significant effect of the starting position with respect to accuracy was observed ($F_{2,92} = 2.11, p = 0.125$).

An investigation of the object orientation with regard to accuracy yields the following observations (Figure S1 b), SI). For the $C_e$ case, there is a clear gradient of accuracy following the increase of orientation difference. Notably, trials of $C_e$ objects with orientation $0^{\circ}$ had an accuracy of 100\%. However, for the other complexity levels, a different pattern can be identified; $90^{\circ}$ was most accurately identified with 94.82\% and 90\% for $C_m$ and $C_h$, respectively. $0^{\circ}$ and $180^{\circ}$ ranked second and third. Interestingly, the object orientation does not have a significant effect on the accuracy ($F_{2,92} = 2.06, p = 0.132$).

Every subject performed 18 trials, and no target object configurations were repeated. Nevertheless, we expected that some improvement in accuracy would begin to appear. Surprisingly, this was not the case; no significant learning effect was observed ($F_{2,92} = 0.88, p = 0.414$), see Figure S1 d), SI.

\subsection{Number of Fixations}

A substantial amount of data acquisition occurs during a solution to this task, as subjects used a minimum of 6 different eye fixations while averaging 92.38 across all trials. The object complexity plays a role in how many fixations are required to solve this task. $C_e$ objects required about 66 fixations, $C_m$ 69 fixations, and $C_h$ 102 fixations on average. The effect is statistically significant ($F_{2,92} = 32.15, p < 0.0001$), see Figure S2 e), SI.

The evaluation of sameness against the number of fixations revealed two major insights (Figure S2 c), SI). Firstly, the same pairings always required significantly more fixations than different pairings ($F_{1,46} = 7.78, p = 0.00761$). Secondly, the same pairings needed at least 10, in some cases up to 20, fixations on average more. Furthermore, error responses required significantly more fixations than correct answers ($F_{1,46} = 9.762, p < 0.003$).

$C_m$ and $C_h$ cases, starting from $P_s$ resulted in the most fixations on average, followed by starting from the $P_c$ and $P_l$ (Figure S2 a), SI). The starting position, similar to the accuracy of answering correctly, does not have any effect with respect to the number of fixations ($F_{2,92} = 1.37, p = 0.258$).  

In terms of object orientation, (Figure S2 b), SI), orientations $0^{\circ}$ and $90^{\circ}$ are similar, varying only a few fixations for the median and upper and lower quartile. In terms of absolute values, a few trials of $C_h$ and orientation of $0^{\circ}$ required about 800 fixations. Notably, these trials started from $P_l$. In summary, larger orientation differences required significantly more fixations regardless of object complexity ($F_{2,92} = 8.31, p = 0.00048$).

Notably, a significant learning effect with respect to the number of fixations is observed ($F_{5,230} = 3.239, p = 0.0075$). This means that participants require fewer fixations (Figure S2 d), SI), hence solving the task more efficiently but not more accurately, as the trials progress.

\subsection{Response Time}

The response time is the time elapsed from the first fixation of the trial to the time when the subject provided the answer. On average, the response time was 47.52s ($\sigma = 30.39$). Among all trials, the shortest response was for an $C_e$ level, starting from $P_s$, with $180^{\circ}$ orientational difference and only taking 4.2s. The longest response time was recorded for a $C_h$ level, starting from $P_l$ and required 298s. 

The complexity of the stimuli affects the response time gradually for $C_e$ (on average 40.03s), and $C_m$ (on average 42.01s) cases, and distinctively for $C_h$ (on average 60.53s) -- increasing object complexity also means a significant increase in response time ($F_{2,92} = 28.87, p < 0.0001$), see Figure S3 e), SI. Furthermore, the response time is approximately a linearly increasing function of the angular difference of the objects; $C_e$ ($n=7, \Delta_t=5.72s$), $C_m$ ($n=10, \Delta_t=4.2s$), and $C_h$ ($n=18, \Delta_t=3.36s$), where $n$ is the number of elements used to create the object and $\Delta_t$ is the normalized response time with respect to a single element (Figure S3 b), SI). This relates well to Sheppard \& Metzler conclusions, but in our 3D active setting.

Similarly to the number of fixations required, the sameness of the stimuli has a distinct effect on the response time ($F_{1,46} = 14.279, p = 0.0004$), see Figure S3 c), SI -- same cases take significantly longer than different ones.

Consistent with other measures, the response time is not significantly affected by the starting position ($F_{2, 92} = 0.12, p = 0.886$), see Figure S3 a), SI. The object orientation, however, does affect response time ($F_{2,92} = 12.95, p < 0.0001$). In general, a lesser orientation difference also means a quicker response time --  $0^{\circ}$ was answered the quickest, followed by $90^{\circ}$, and $180^{\circ}$, see Figure S3 b), SI.

Subjects seem to develop more efficient strategies with increasing trials completed. Starting at about 47s (Mdn.) at the first trial, the response time drops to about 34s (Mdn.) for trials two to four and drops further to 29s Mdn. at trials five and six. For $C_h$ cases, a drop from the first trial (70s Mdn.) to the second trial (about 50s Mdn.) can be seen (Figure S3 d), SI). Overall, looking at the impact of progressing trials and their response time, a significant effect is noticed ($F_{5,230} = 6.01, p = 0.0003$).

\subsection{Movement}

The mean of head movement was 16.62m. The amount of head movement slightly increased from complexity cases of $C_e$ to $C_m$ but increased more distinctly for $C_h$ cases -- the object complexity significantly affects the amount of head movement ($F_{2,92} = 35.35, p < 0.0001$, Figure S4 d), SI). 

Aligned with the number of fixations, response time and accuracy, the amount of movement is greater for the same object pairings across all complexity levels ($F_{1,46} = 31.37, p < 0.0001$), see Figure S4 c), SI. For different cases, the increased upper and lower quartiles indicate that more uncertainty across different subjects in how to approach this case was involved.

We found no relationship between amount of head movement and starting position (Figure S4 a), SI). However, there exists a significant effect on the correctness of the answer and head movement (Figure S4 e), SI). Error responses were accompanied by significantly more head movement ($F_{1,46} = 41.56, p < 0.0001$).

A clear trend ($F_{2,92} = 22.74, p < 0.0001$) can be observed between the amount of head movement and the amount of orientational difference (see Figure S4 b), SI); at $0^{\circ}$ the least amount of movement was required, at $90^{\circ}$ an increase of 2-5m on average is recorded, and at $180^{\circ}$ an additional increase of 1-5m across all complexity classes is recorded.

A significant reduction in head movement is noticable over the course of the trials ($F_{5,230} = 5.403, p = 0.0001$). This means that participants show a learning effect in the sense that they execute a strategy with less head moment as the experiment progresses. For $C_m$ and $C_h$ cases, a trend is not visible directly, but for the $C_e$ cases, it is (Figure S4 d), SI). $C_h$ cases start off at the first trial with just above 20m and drop to the absolute mean value of 16.62m and stay steady, marginally falling below and exceeding it repetitively; similarly, for the $C_m$ case, where no learning trend can be observed. However, the $C_e$ case, while noticing a slight up-trend for the second trial, consecutively decreases from about 16m down to about 10m, which is the equivalent of an improvement of 37.5\%.

Lastly, in Figure S5, SI, we plot normalized measured variables (accuracy, amount of fixations, response time, and movement) against trial number. It is easily seen that every variable improves over the trials except for accuracy. 

\subsection{Fixation Patterns}

Lastly, we looked at fixation patterns. We considered the ratio of fixations landing on each object respectively and fixation groupings, specifically fixations falling on the same object before shifting away.

To evaluate the fixation ratio, we looked at the total number of fixations for either object. The object with the most fixations is considered the primary object, and the object with fewer or the same number of fixations is the secondary object. On average, the primary object accounted for $59.53\%$ of fixations, and the secondary object for $40.47\%$.

None of the experimental variables have a significant effect on the fixation ratio: object complexity ($F_{2,92} = 0.844, p = 0.433$), starting position ($F_{2,92} = 0.83, p = 0.438$), object orientation ($F_{2,92} = 1.32, p = 0.271$), object sameness ($F_{2,46} = 0.19, p = 0.664$), trial progression ($F_{5,230} = 1.08, p = 0.368$), and correctness of answer ($F_{1,46} = 1.421, p = 0.239$) (Figure S6, SI). 

However, for some configurations, a difference in fixation ratio is noticeable. For instance, the difference in the average number of fixations with respect to object orientation decreases with increasing orientational difference (Figure S6 c, SI). 

Next, we looked at fixation groupings, which means the number of fixations on one object before changing focus to the other object (Figure S7, SI). On average, $18.7\%$ ($\sigma = 10.01\%$) are fixations that change focus between each object every time -- here, we call them single fixations. The remaining $81.3\%$ of fixations are divided as follows: couple fixations $18.43\%$ ($\sigma = 11.48\%$), triple fixations $12.26\%$ ($\sigma = 8.51\%$), quadruple fixations $8.57\%$ ($\sigma = 7.41\%$), quintuple fixations $6.71\%$ ($\sigma = 7.20\%$), sextuple fixations $4.87\%$ ($\sigma = 7.47\%$), septuple fixations $3.74\%$ ($\sigma = 5.94\%$), octuple fixations $3.23\%$ ($\sigma = 6.13\%$), and higher groupings ($>8$) $18.41\%$ ($\sigma = 17.90\%$).

The object complexity has an significant effect on single ($F_{2,92} = 8.53, p = 0.0003$), couple ($F_{2,92} = 9.35, p = 0.0002$), triple ($F_{2,92} = 4.34, p = 0.015$), quintuple ($F_{2,92} = 6.428, p = 0.002$), octuple ($F_{2,92} = 5.10, p = 0.007$) and higher ($F_{2,92} = 15.56, p < 0.0001$) fixation groupings (Figure S7 a, SI). Notably, for single, couple, and triple, the probability of occurrence significantly decreases with increasing object complexity. While for octuple and higher groupings, the opposite is true -- their occurrence increases with increasing object complexity.

While object complexity has a significant effect on grouping patterns, the starting position does not. No investigated fixation group sizes showed to be significantly affected by the starting position: single ($F_{2,92} = 0.053, p = 0.948$), couple ($F_{2,92} = 0.884, p = 0.416$), triple ($F_{2,92} = 0.717, p = 0.49$), quadruple ($F_{2,92} = 1.01, p = 0.366$), quintuple ($F_{2,92} = 0.24, p = 0.78$), sextuple ($F_{2,92} = 0.17, p = 0.843$), septuple ($F_{2,92} = 0.61, p = 0.541$), octuple ($F_{2,92} = 1.22, p = 0.299$), higher ($F_{2,92} = 1.39, p = 0.25$).

However, the object orientation has a significant effect on single ($F_{2,92} = 8.97, p = 0.0002$) and couple ($F_{2,92} = 15.436, p < 0.0001$) groupings as this is the dominant method for object orientation $0^{\circ}$ and decreases steadily with increasing object orientation. Similar to object complexity, larger fixation groupings are affected by object orientation as well. Specifically, septuple ($F_{2,92} = 3.305, p = 0.041$) and higher ($F_{2,92} = 7.44, p = 0.001$) groupings occur more frequently with increasing object orientation (Figure S7 c, SI).

The sameness of the object had largely no significant effect on fixation groups (Figure S7 d, SI). Only single ($F_{2,92} = 5.25, p = 0.02$) and septuple ($F_{2,46} = 13.7, p = 0.0005$) groups are more significantly used for the same objects than different ones. As trials advanced, subjects used single groupings progressively less ($F_{2,92} = 1.33, p = 0.25$). A similar trend is observed for couple, triple, quadruple and quintuple groups -- none are significant, however. Larger groupings see an increase in probability as trials proceed. Notably, only a few sparse data points are recorded for octuple pairings up to trial 8. Octuple pairings are more regularly seen for trials 9-19 (Figure S7 e, SI).

The correctness of the answer significantly correlated with single ($F_{2,92} = 3.173, p = 0.081$) and septuple ($F_{1,46} = 10.624, p = 0.002$) pairs. Single and septuple fixation groups are significantly used more for correct answers than error responses (Figure S7 f, SI).

The fixation groupings reveal a purposeful gaze toward the solution of 3D visuospatial problems. Further analysis and experiments are hoped to flesh out behaviours comprising human functional vision.

\section{Discussion}

The goal of this study was to examine functional vision in human subjects, specifically, how they solve a visuospatial problem in a three-dimensional space.

We addressed this by developing a three-dimensional version of the well-known same-different task as a probe and an experimental setup allowing natural, visual problem-solving and precision recording. Such physical object comparisons seem a fundamental cognitive ability \cite{carroll1993human}.

Our main results follow. People are very good at this task, even in difficult cases. No training trials were required. The range of response times from simplest to most complex cases ranges from 4 to 298 sec. and accuracy from 80\% to 100\%. A great deal of data acquisition is occurs during all trials with the range of eye movements (separate fixations and separate images processed) from 6 to 800 fixations. 

Furthermore, we showed that not only multiple fixations are required, but also multiple fixations in sequence on the same stimulus. Only about $20\%$ of all fixations are single fixations, and fixation groups get larger and more frequent with increasing levels of complexity and orientation. These groups seem to develop throughout the course of the trial. Simpler groupings (single, couple, triple) are replaced by more complex ones (septuple, octuple, larger) as the trial progresses. This hints that subjects use what they know to dynamically compose visuospatial strategies. In our analysis of fixation ratios, subjects did not simply observe each object with the same number of fixations. They chose one object as their primary object ($59.53\%$ of total fixations, regardless of experimental setup) and spent just about $40\%$ of fixations on the secondary to solve this problem -- brute-force approaches would have averaged a $50:50$ ratio leading to the conclusion that subjects did not use random or uninformed search strategies. Could they be building internal models of the objects, which are then compared? This possibility needs further investigation. In Figure S8, SI, we illustrate a set of fixations observed in a trial with $C_m$ complexity, same objects, starting from $P_c$, presented at $90^{\circ}$. This pattern looks like an observation strategy; the gaze goes back and forth between stimuli. Interestingly, the fixations land on similar-looking areas. It appears that the subject compares object features. 

No statistical change was observed in accuracy with increasing trials for individual subjects. However, a change was observed in the number of fixations, response time, and amount of head movement. This is surprising, as the accuracy did not significantly change throughout the trials and shows that the set of visuospatial problem-solving techniques, innate and learned through a lifetime, generalized well to this specific task. However, the decrease in the number of fixations, response time, and amount of head movement shows that participants may have fine-tuned them for efficiency. This learning effect seems counterintuitive, especially when compared to modern computational attempts at active learning (however, see \cite{Taylor2021} for review and a promising change). 

Our work also shows consistency with the classic version of the same-different task \cite{Shepard2019} in that the time required to determine if two perspective drawings portray objects of the same three-dimensional shape is found to be a linearly increasing function of the angular difference in the portrayed orientations for the two objects.

Other three-dimensional stimuli were considered. Some examples are the well-studied greebles families \cite{Gauthier1997}, the \textit{CLEVR} data set \cite{Johnson2017a}, or \textit{T-LESS} \cite{Hodan2017}. All are virtual but could be physically created, for instance, using fused manufacturing modelling. Various versions of the greebles objects have been introduced. These objects would function well as a stimulus for this experiment as they are textureless, and a common-coordinate system could be defined easily (greebles are all structured similarly). However, different greebles appear quite different and would make the same-different task trivial. The \textit{CLEVR} data set does use simple blocks to build the stimulus, similar to \textit{TEOS}; however, neither a systematic measure for self-occlusion nor a common coordinate system to define the object pose exists. \textit{T-LESS} objects do have an associated pose and also have a textureless appearance, but the objects are easily differentiable. \textit{TEOS} combines crucial properties to discover any patterns in solving the same-different task; novel and unfamiliar, textureless appearance, known complexity, common coordinate system, and varying self-occlusion.

Humans use vision for a vast array of behaviours in the real world; visuospatial intelligence is much more than simply detecting a stimulus or recognizing an object or scene. Unfortunately, past methodologies have limited studies beyond these, and the fundamental questions of three-dimensional visuospatial intelligence in the real world remain. Where do we look? How do we look? How do we move? How do we seek out the data that enables problems to be solved? The first steps towards these answers are presented along with an experimental infrastructure appropriate for many further studies. Our data shows that we actually do a great deal of 'looking' in order to solve a problem. Adult humans do not need to learn where to look and exhibit an array of complex fixation patterns. We also move about as we look, most likely because we choose what we wish to see and from what vantage in order to support the task at hand. Although the use of our eyes may seem effortless, the complexity of actions is staggering and unravelling their purpose - the 'why' behind all this looking - poses an exciting challenge.

\section{Funding}\label{sec:si}

This material is based upon work supported by the Air Force Office of Scientific Research under award numbers FA9550-18-1-0054 and  FA9550-22-1-0538 (Computational Cognition and Machine Intelligence, and Cognitive and Computational Neuroscience portfolios); the Canada Research Chairs Program (grant number 950-231659); Natural Sciences and Engineering Research Council of Canada (grant numbers RGPIN-2016-05352 and RGPIN-2022-04606).

\bibliographystyle{IEEEtran}
\bibliography{library}

\newpage
\onecolumn
\section*{Supporting Information}

\renewcommand{\thefigure}{S1}
\begin{figure}[h]
     \centering
     \begin{subfigure}[b]{0.49\linewidth}
         \centering
         \includegraphics[width=\linewidth]{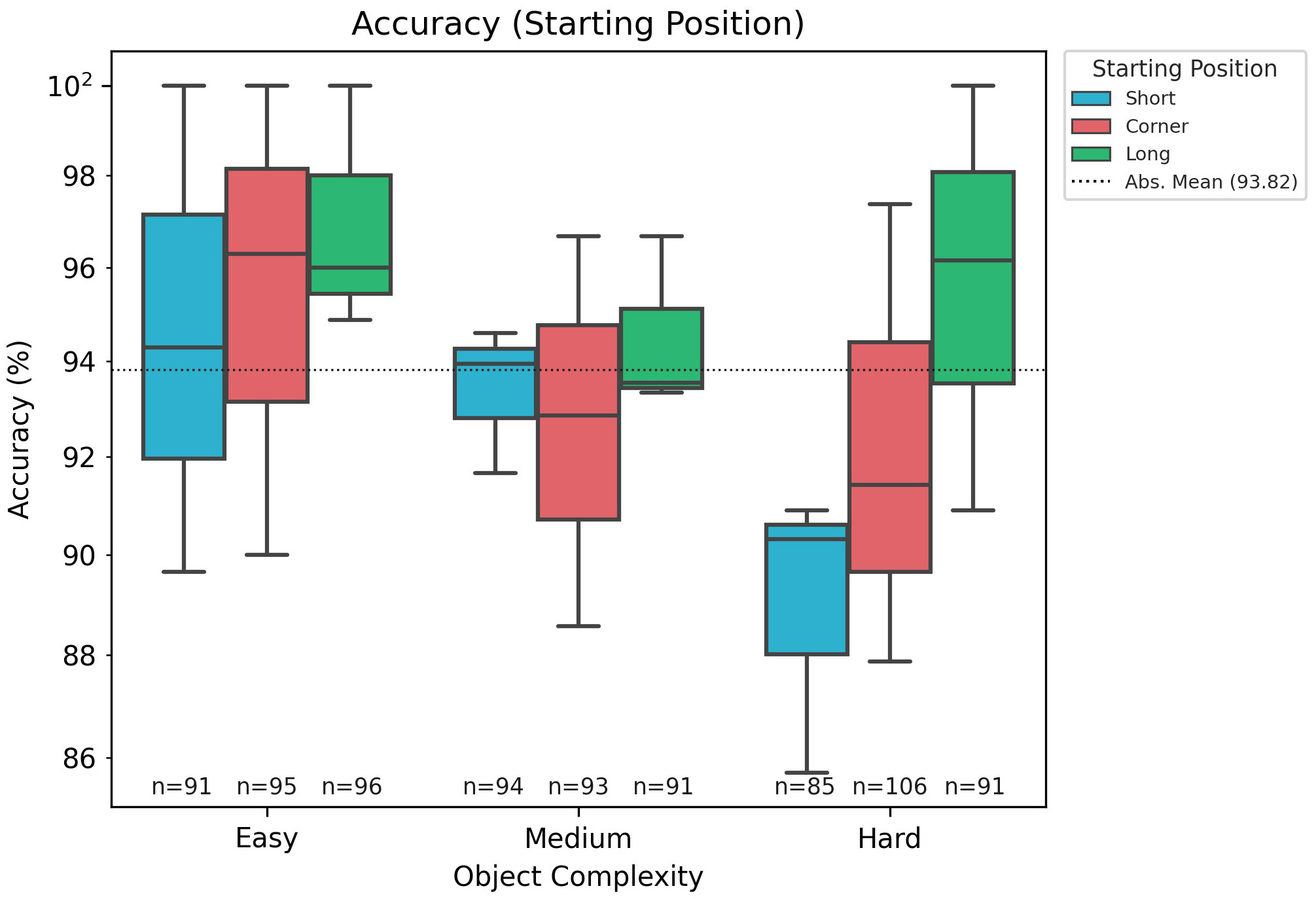}
         \caption{}
         \label{fig:res_acc_start}
     \end{subfigure}
     \hfill
     \begin{subfigure}[b]{0.49\linewidth}
         \centering
         \includegraphics[width=\linewidth]{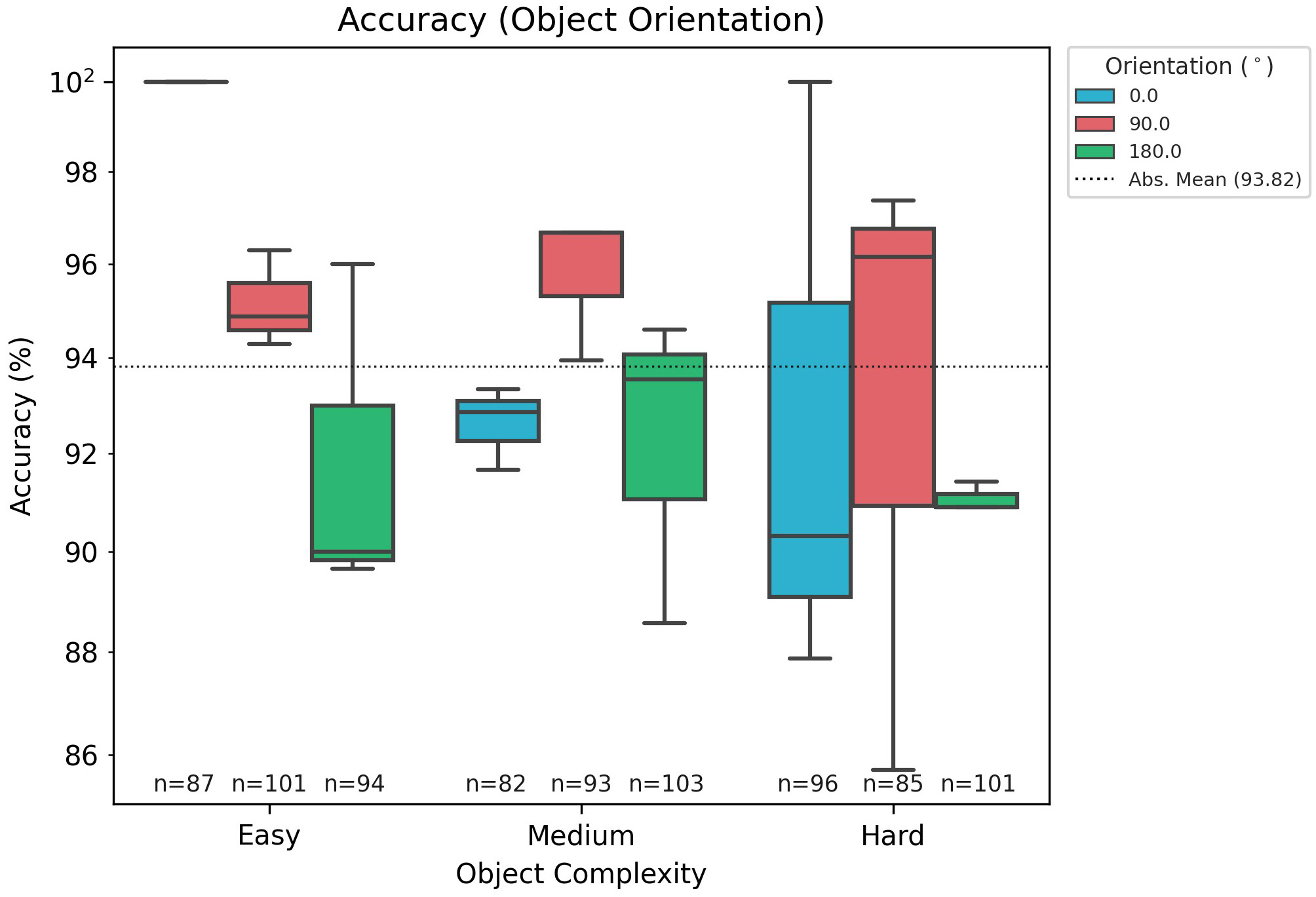}
         \caption{}
         \label{fig:res_acc_orie}
     \end{subfigure}
     \hfill
     \begin{subfigure}[b]{0.49\linewidth}
         \centering
         \includegraphics[width=\linewidth]{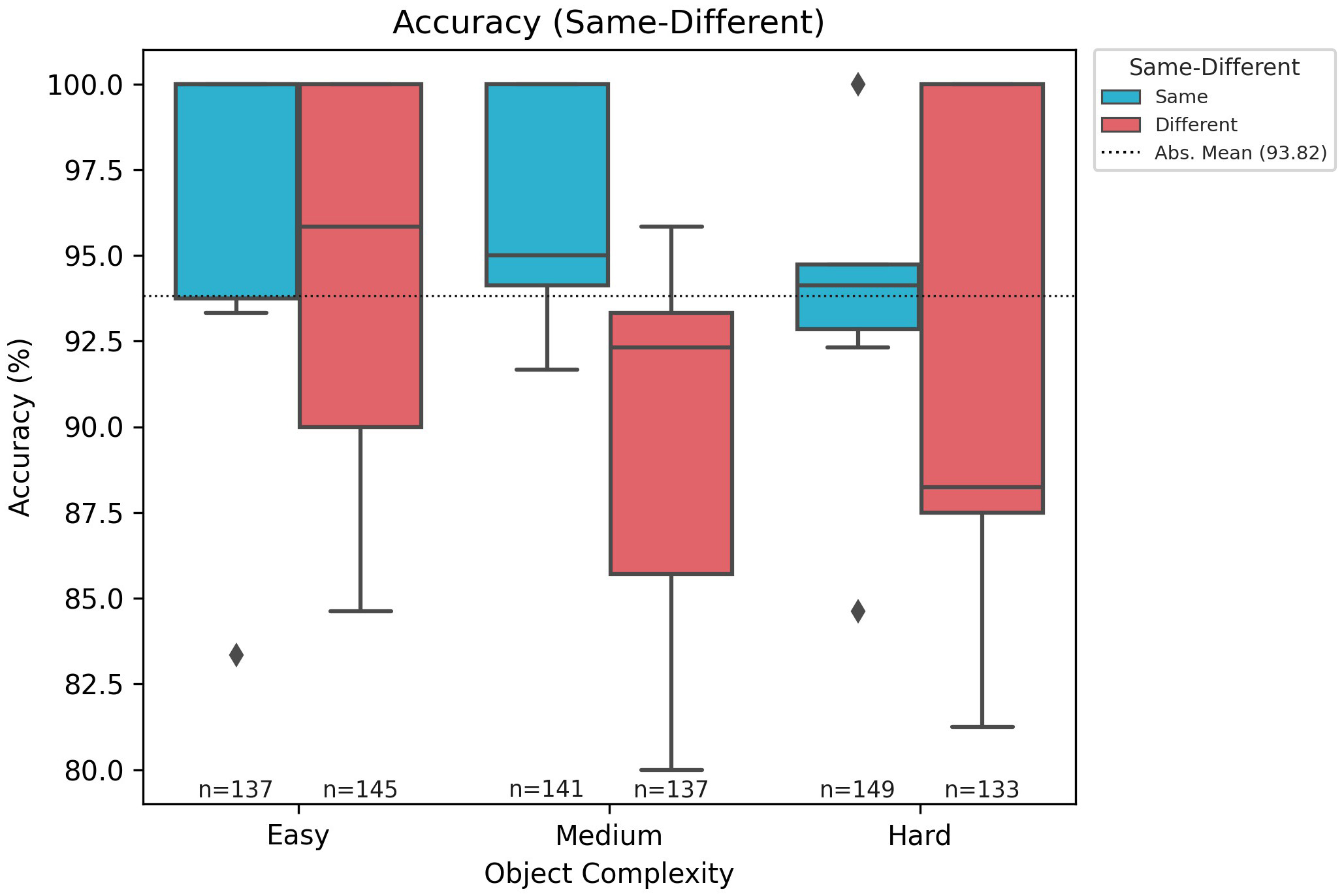}
         \caption{}
         \label{fig:res_acc_sd}
     \end{subfigure}
     \hfill
     \begin{subfigure}[b]{0.49\linewidth}
         \centering
         \includegraphics[width=\linewidth]{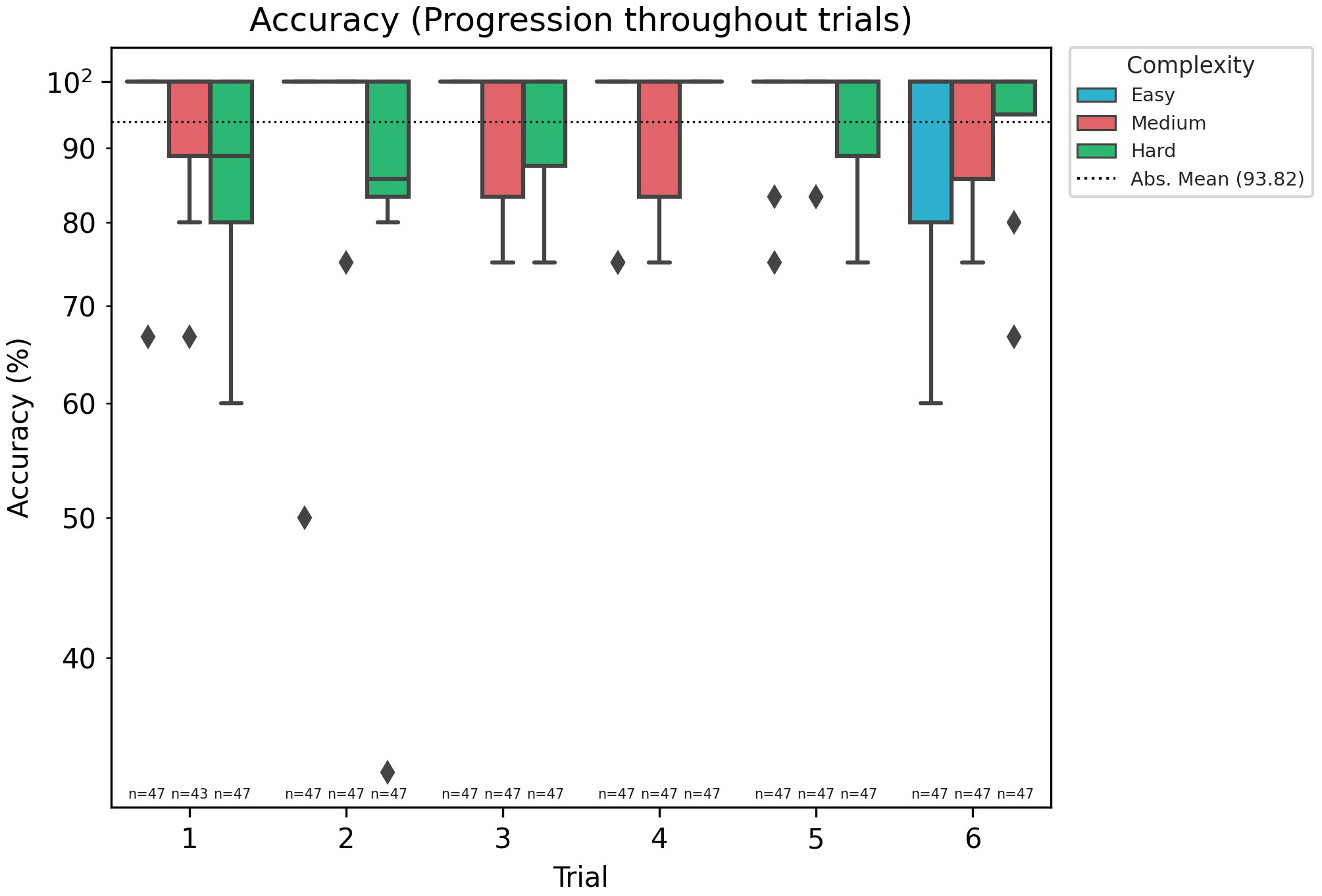}
         \caption{}
         \label{fig:res_acc_prog}
     \end{subfigure}
     \caption{Plots illustrating the accuracy measured against different experimental variables. \textbf{(a)} The effect of the starting position across the different object complexities. The best accuracy for $C_e$ (easy) is from $P_c$. For $C_m$ the highest accuracy was achieved starting from $P_s$. Lastly, $C_h$ was most accurately answered starting from $P_l$, meaning that every complexity class had its unique starting position from which it performed best on average.
    \textbf{(b)} The effect of object orientation. For $C_e$ cases, a clear gradient of accuracy following the increase of orientation is visible. Notably, all trials with $C_e$ with $0^{\circ}$ orientation were always answered correctly. This is the only case where this is true. However, for the other complexity levels, a different pattern can be identified; $90^{\circ}$ was most accurately identified with 94.82\% and 90\% for $C_m$ and $C_h$, respectively. $0^{\circ}$ and $180^{\circ}$ ranked second and third. Interestingly, the object orientation does not have a significant effect on the accuracy. \textbf{(c)} The effect of object sameness. Accuracy is not significantly affected. If the objects are the same; generally, a higher accuracy is seen (94.3\%) than for different pairings (91.6\%). Lastly, the progression throughout trials \textbf{(d)} revealed that no improvement in accuracy appeared --  no significant learning effect was observed.}
\end{figure} 

\renewcommand{\thefigure}{S2}
\begin{figure}
     \centering
     \begin{subfigure}[b]{0.49\linewidth}
         \centering
         \includegraphics[width=\linewidth]{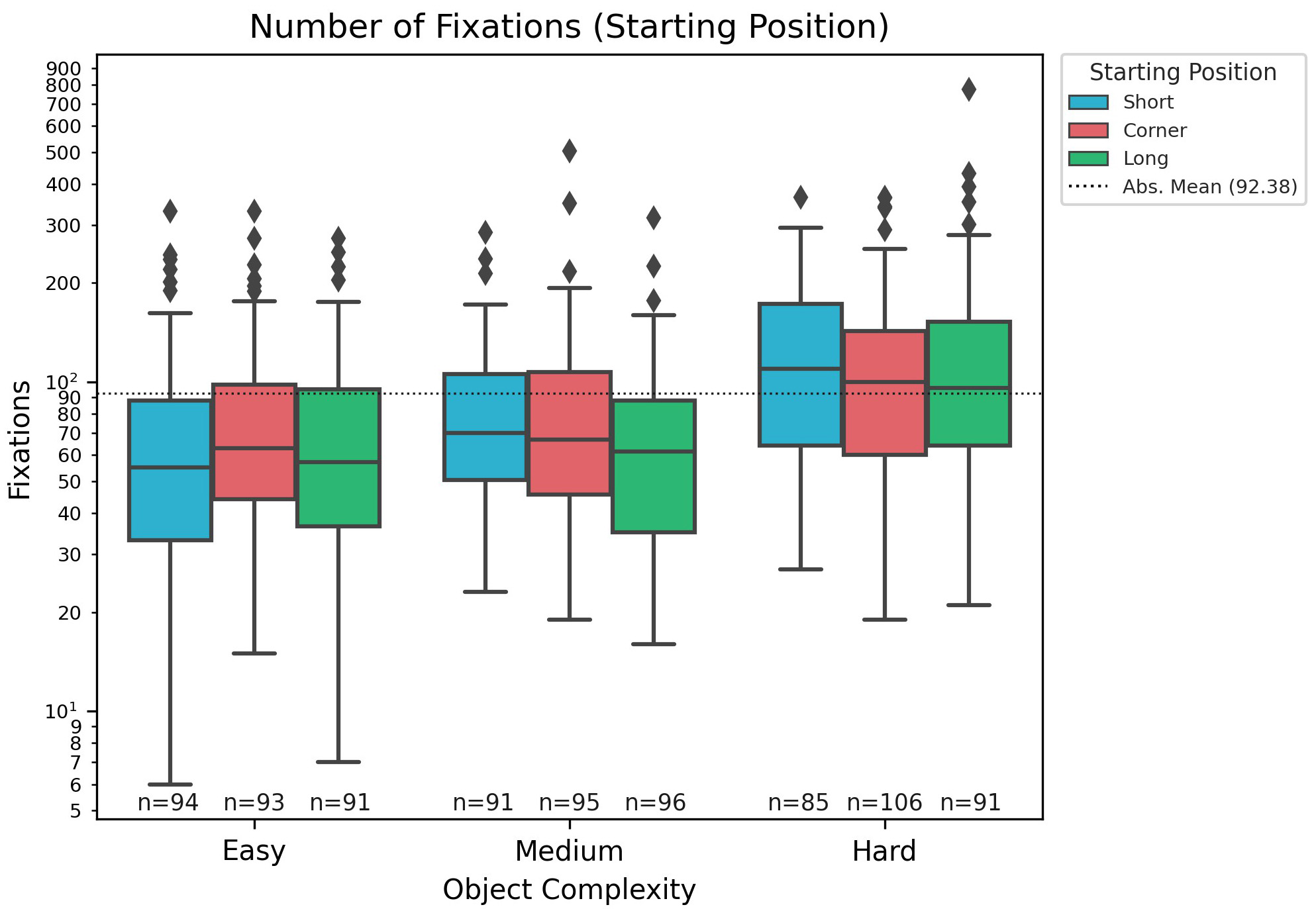}
         \caption{}
         \label{fig:res_fix_pos}
     \end{subfigure}
     \hfill
     \begin{subfigure}[b]{0.49\linewidth}
         \centering
         \includegraphics[width=\linewidth]{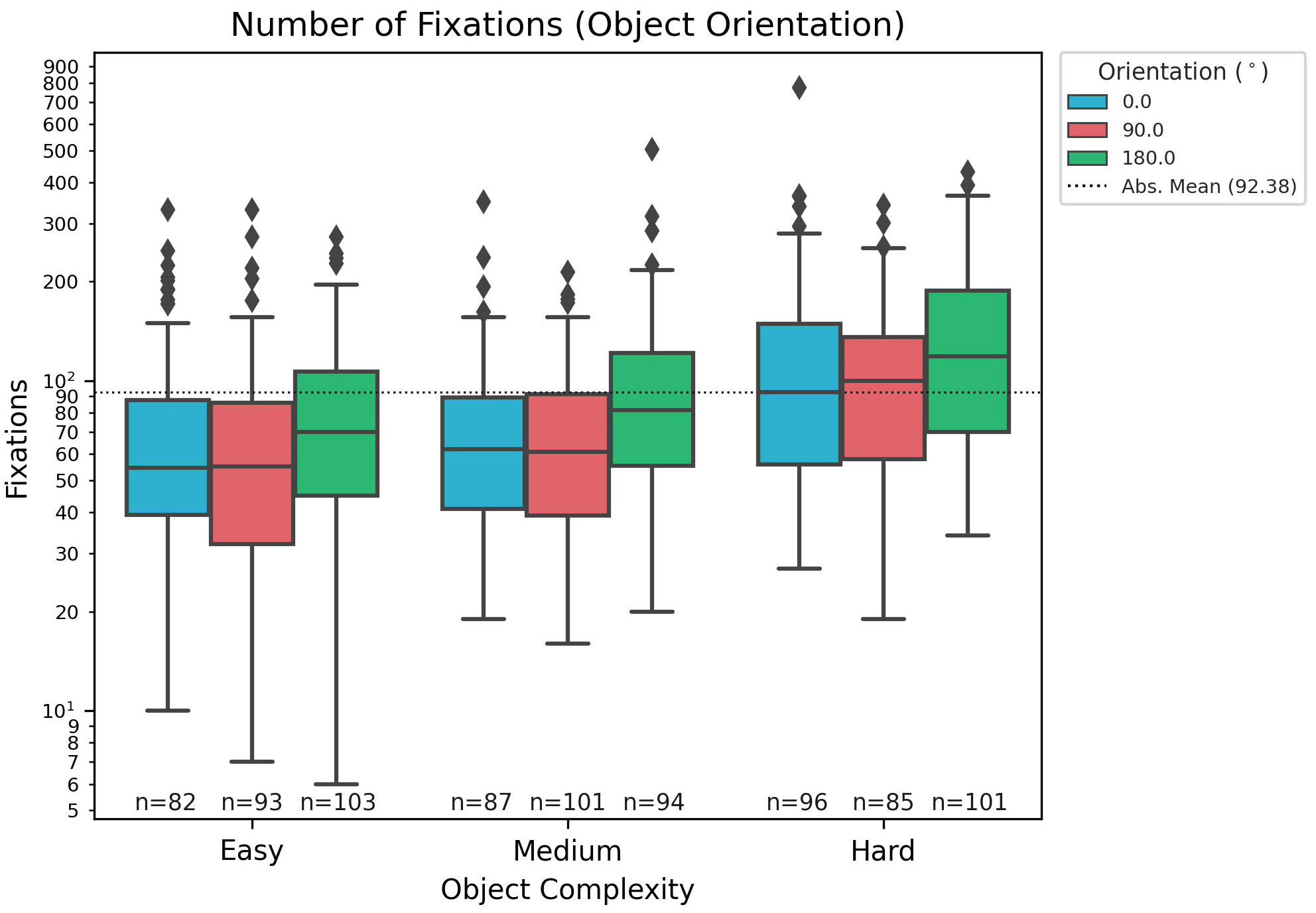}
         \caption{}
         \label{fig:res_fix_orien}
     \end{subfigure}
     \hfill
     \begin{subfigure}[b]{0.49\linewidth}
         \centering
         \includegraphics[width=\linewidth]{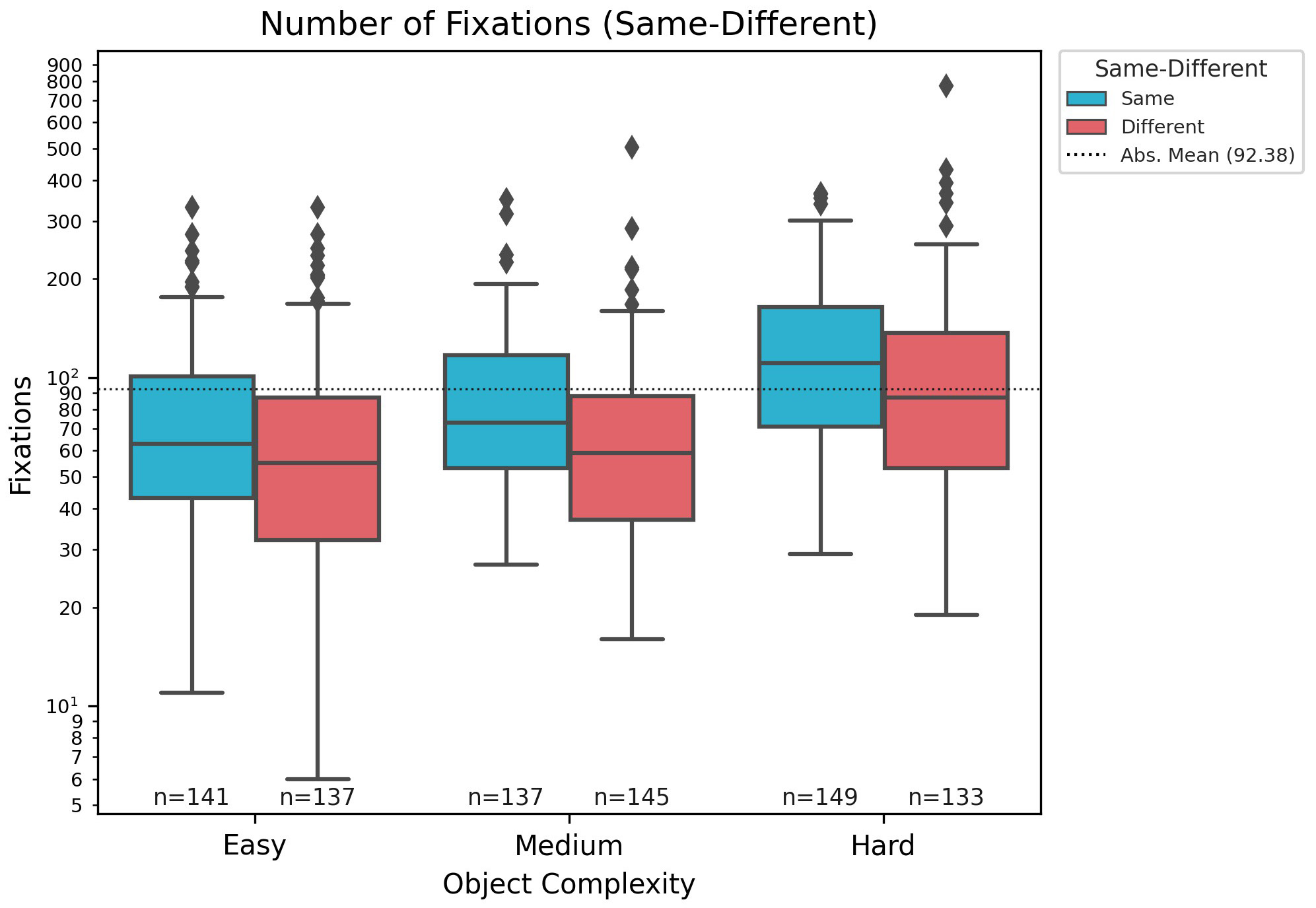}
         \caption{}
         \label{fig:res_fix_same}
     \end{subfigure}
     \hfill
     \begin{subfigure}[b]{0.49\linewidth}
         \centering
         \includegraphics[width=\linewidth]{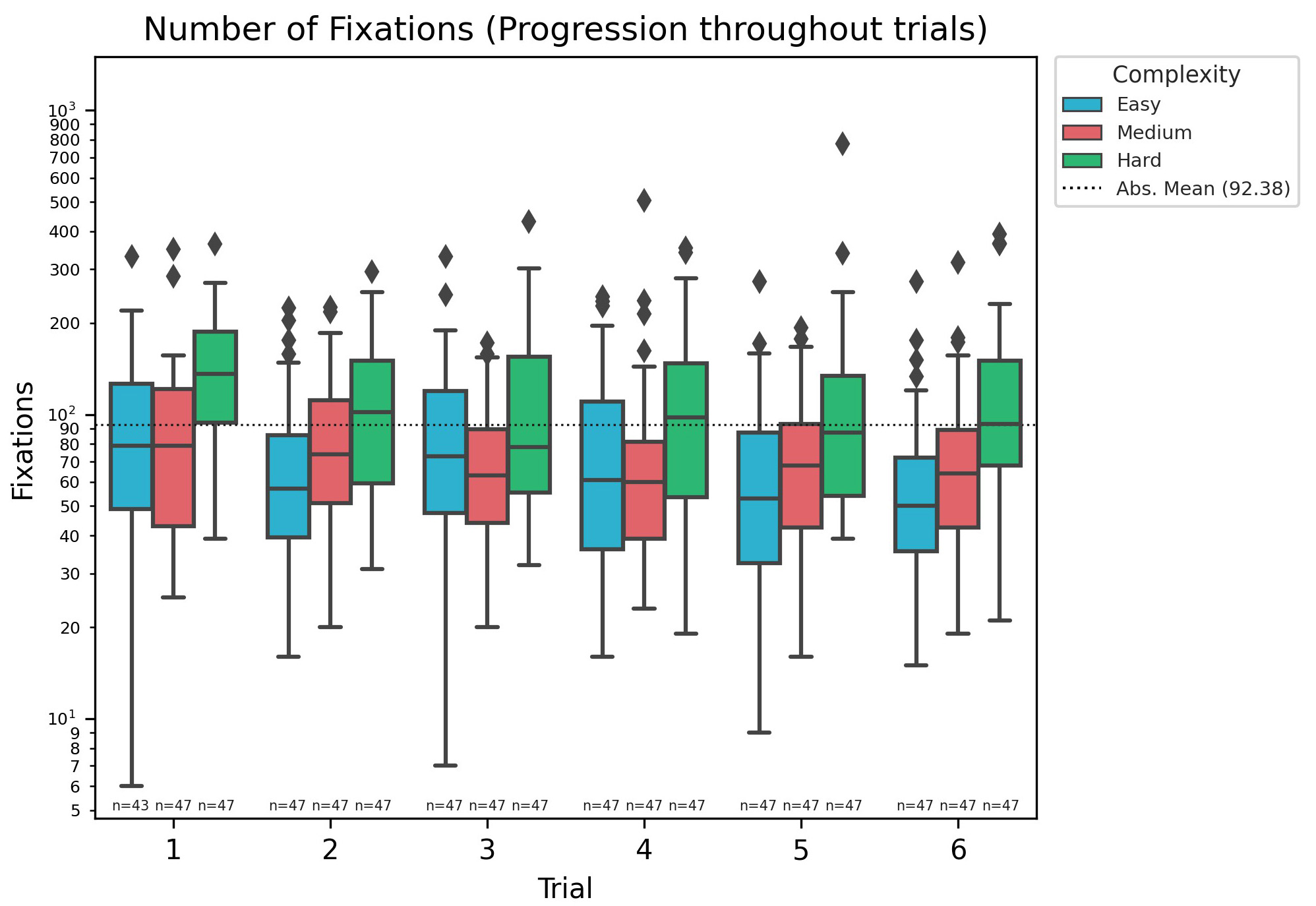}
         \caption{}
         \label{fig:res_fix_prog}
     \end{subfigure}
     \hfill
     \begin{subfigure}[b]{0.49\linewidth}
         \centering
         \includegraphics[width=\linewidth]{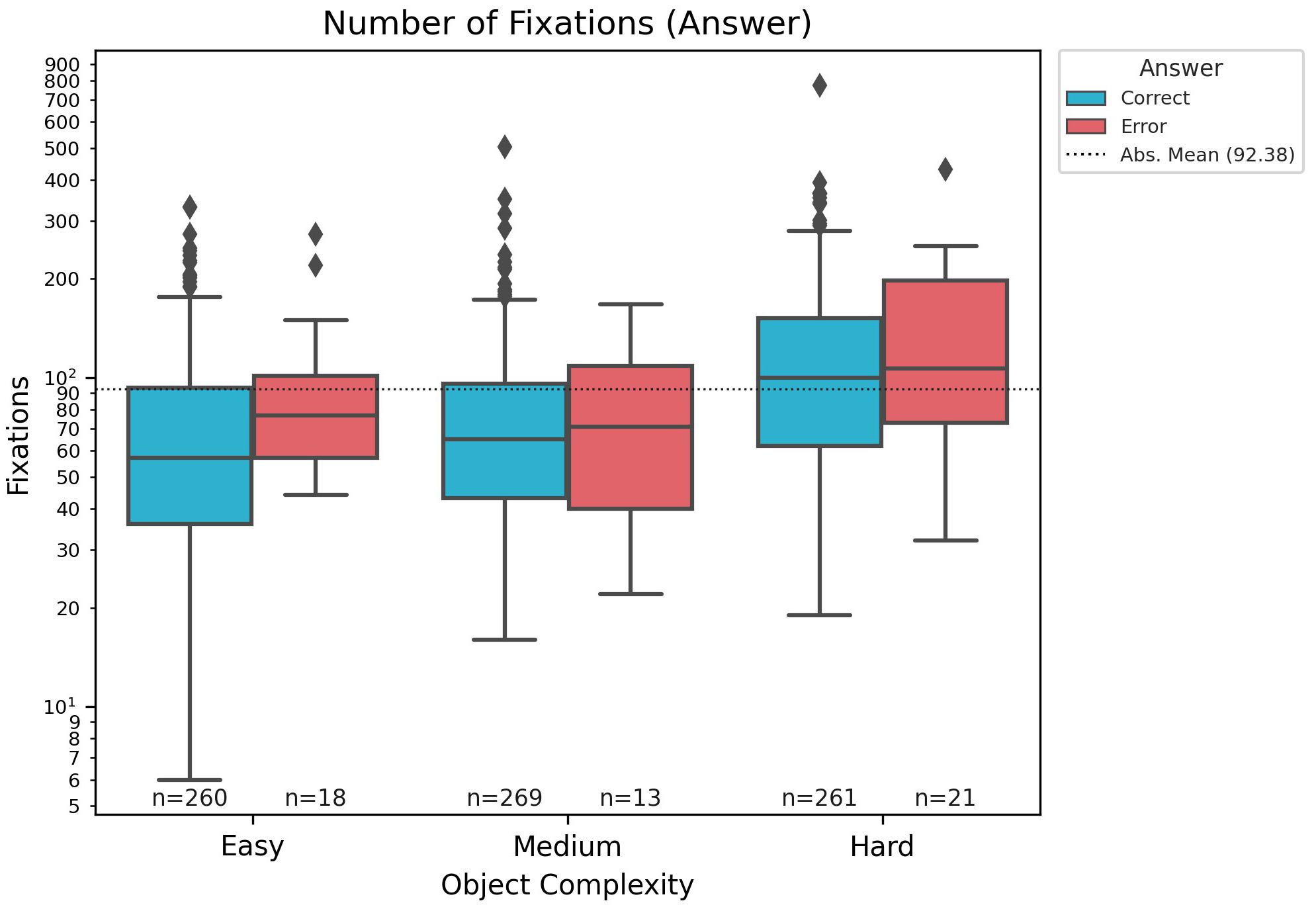}
         \caption{}
         \label{fig:res_fix_answer}
     \end{subfigure}
     \caption{The number of fixations against different experimental variables. \textbf{(a)} The effect of the starting position. For $C_m$ and $C_h$ cases, starting from $P_s$ resulted in the most fixations on average, followed by starting from the $P_c$ and $P_l$. \textbf{(b)} The effect of object orientation. Orientations $0^{\circ}$ and $90^{\circ}$ are similar, varying only a few fixations for the median and upper and lower quartile. In terms of absolute values, a few trials of $C_h$ and orientation of $0^{\circ}$ required about 800 fixations. Notably, these trials started from $P_l$. Larger orientation differences required significantly more fixations regardless of object complexity. \textbf{(c)} The evaluation of sameness against the number of fixations. The same pairings always required significantly more fixations than different pairings. The same pairings needed at least 10, in some cases up to 20, fixations on average more. Additionally, error responses required significantly more fixations than correct answers. \textbf{(d)} A significant learning effect with respect to the number of fixations is observed. \textbf{(e)} The effect of correctness. Error responses result in more fixations than correct answers.}
     \label{fig:res_fix_all}
\end{figure}

\renewcommand{\thefigure}{S3}
\begin{figure}
     \centering
     \begin{subfigure}[b]{0.49\linewidth}
         \centering
         \includegraphics[width=\linewidth]{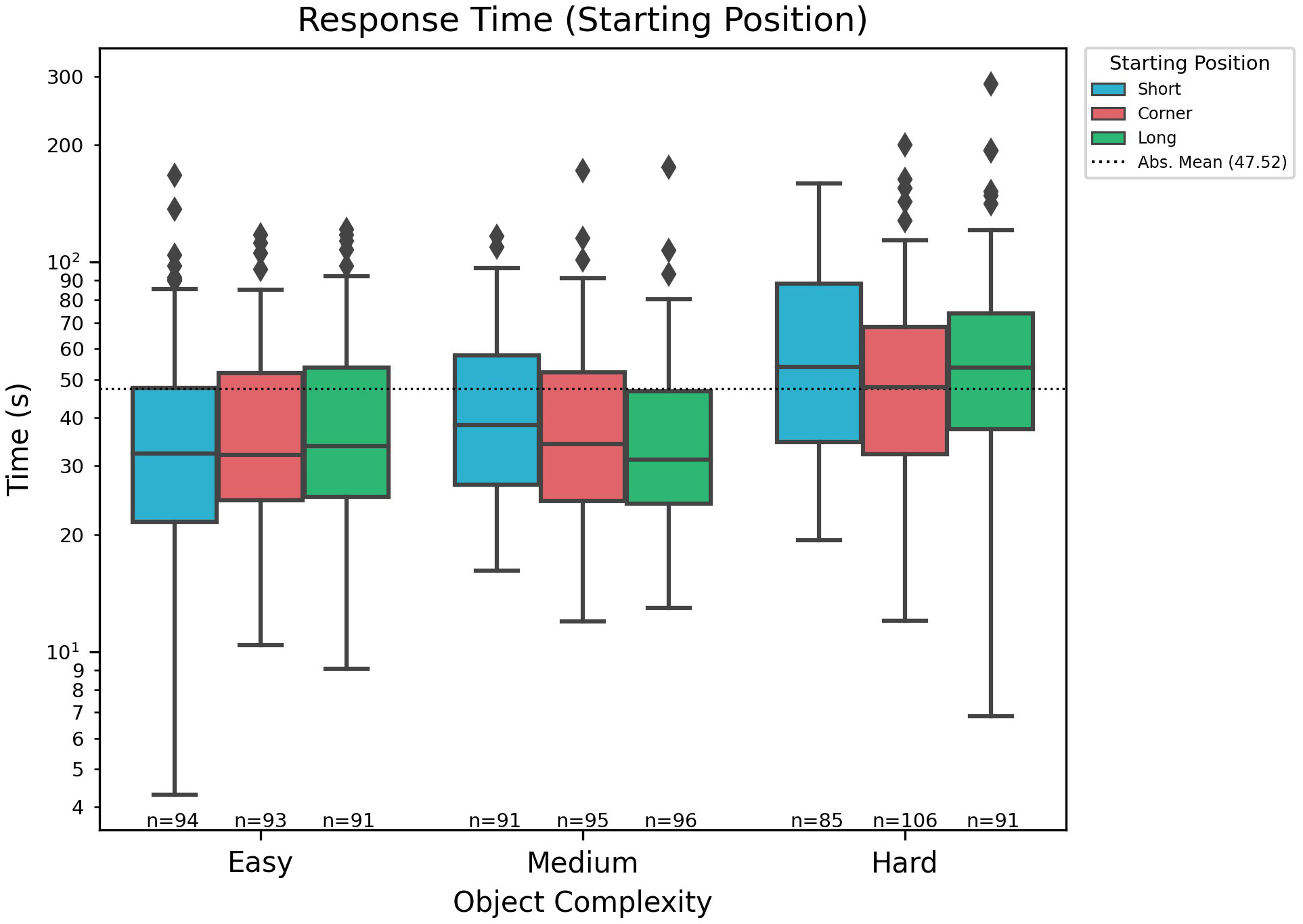}
         \caption{}
         \label{fig:res_respo_pos}
     \end{subfigure}
     \hfill
     \begin{subfigure}[b]{0.49\linewidth}
         \centering
         \includegraphics[width=\linewidth]{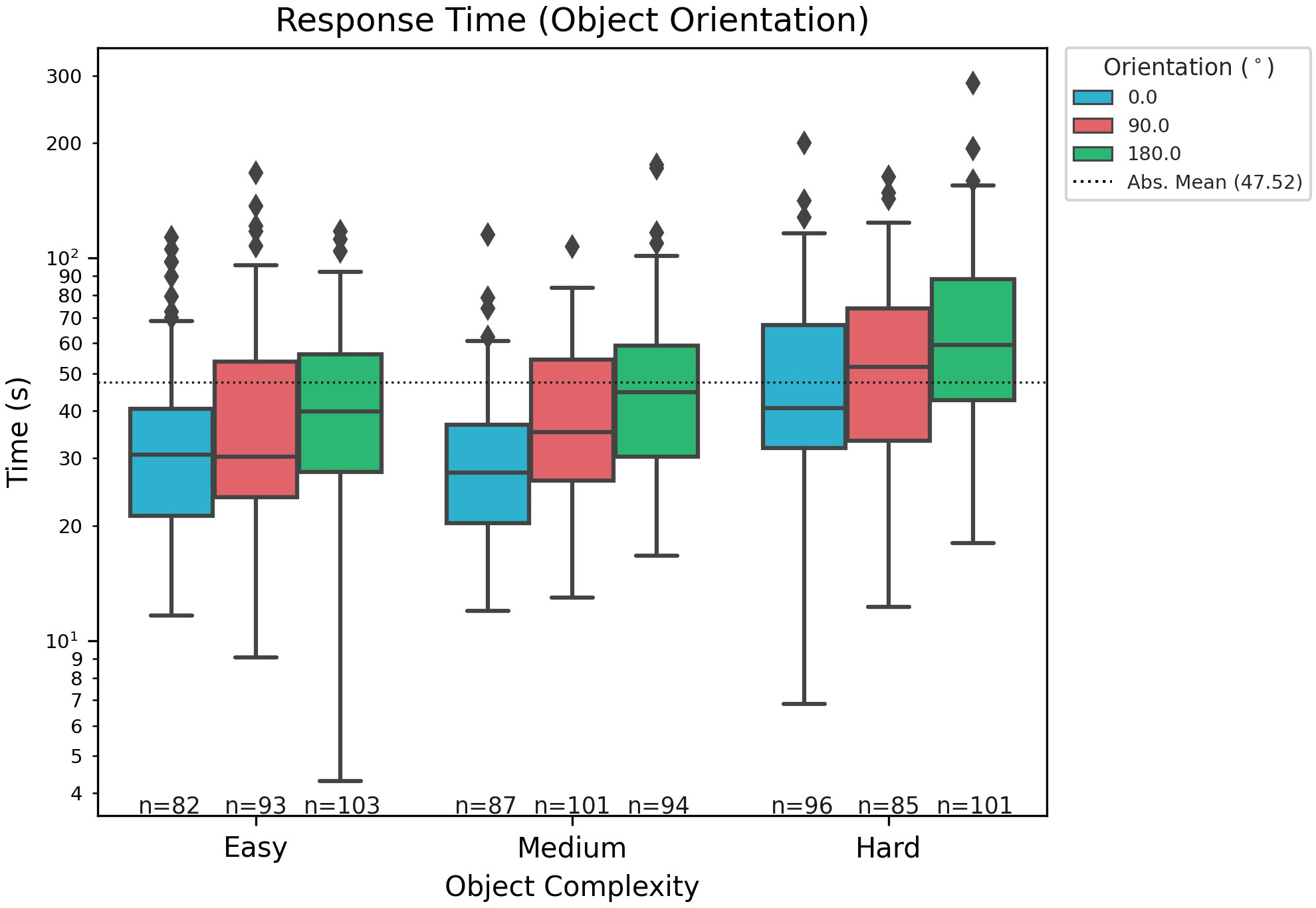}
         \caption{}
         \label{fig:res_respo_orien}
     \end{subfigure}
     \hfill
     \begin{subfigure}[b]{0.49\linewidth}
         \centering
         \includegraphics[width=\linewidth]{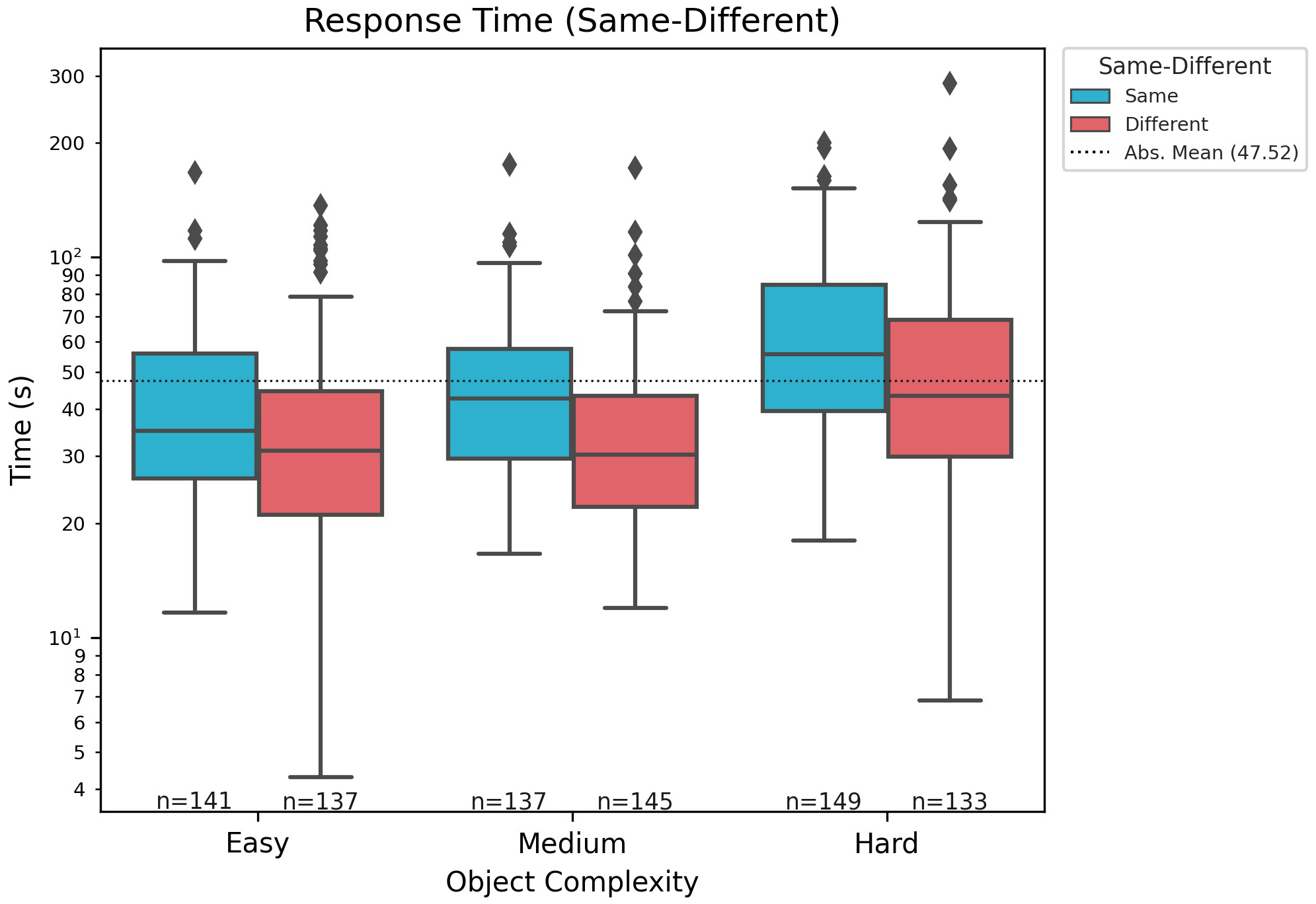}
         \caption{}
         \label{fig:res_respo_same}
     \end{subfigure}
     \hfill
     \begin{subfigure}[b]{0.49\linewidth}
         \centering
         \includegraphics[width=\linewidth]{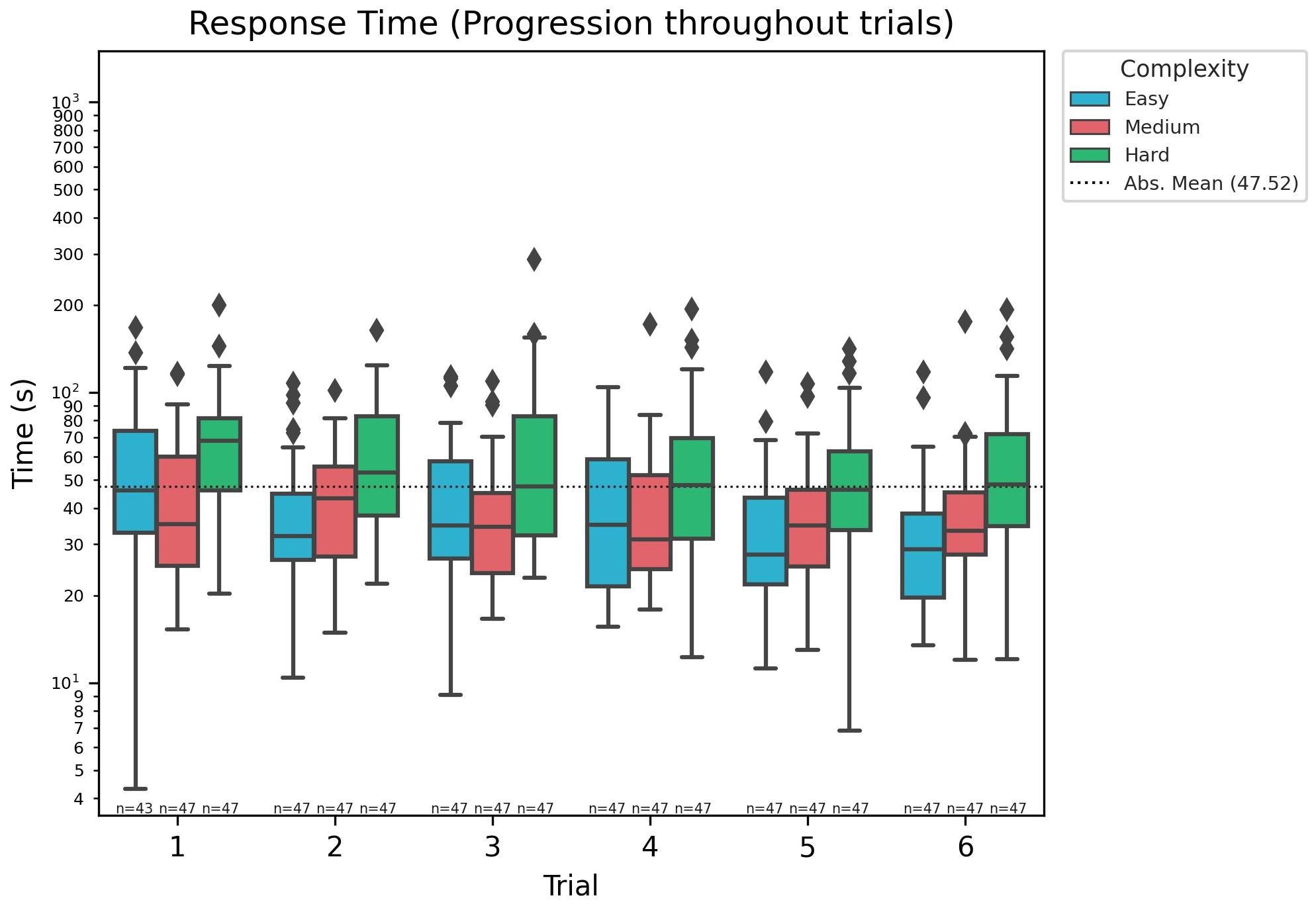}
         \caption{}
         \label{fig:res_respo_progr}
     \end{subfigure}
     \hfill
     \begin{subfigure}[b]{0.49\linewidth}
         \centering
         \includegraphics[width=\linewidth]{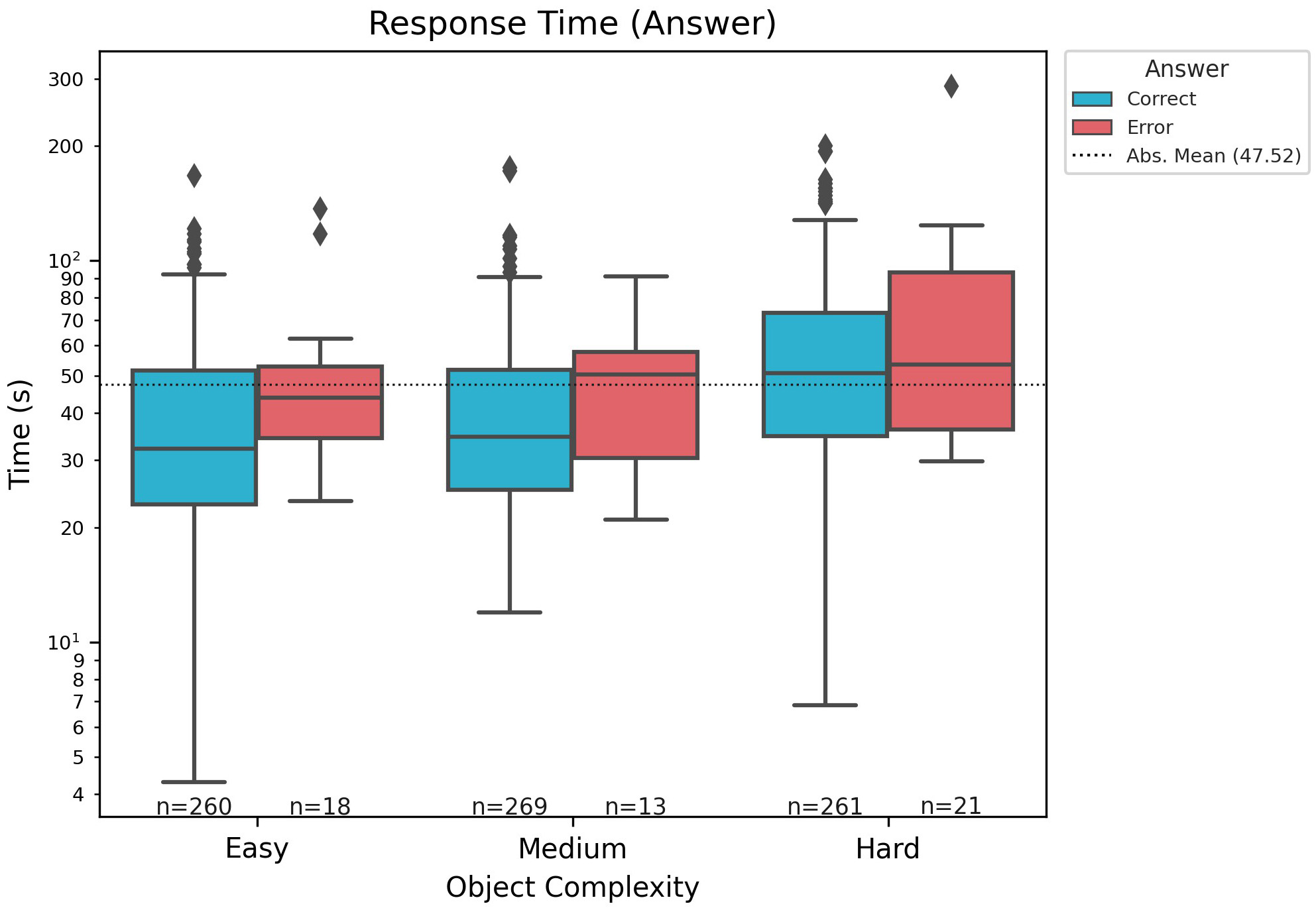}
         \caption{}
         \label{fig:res_respo_answer}
     \end{subfigure}
     \caption{Response Time against different experimental variables. \textbf{(a)} Consistent with other measures, the response time is not significantly affected by the starting position. \textbf{(b)} The object orientation, however, does affect response time. In general, a lesser orientation difference also means a quicker response time --  $0^{\circ}$ was answered the quickest, followed by $90^{\circ}$, and $180^{\circ}$. \textbf{(c)} The sameness of the stimuli has a distinct effect on the response time -- same cases take significantly longer than different ones. \textbf{(d)} Subjects seem to develop more efficient strategies with increasing trials completed. Starting at about 47s (Mdn.) at the first trial, the response time drops to about 34s (Mdn.) for trials two to four and drops further to 29s Mdn. at trials five and six. For $C_h$ cases, a drop from the first trial (70s Mdn.) to the second trial (about 50s Mdn.) can be seen. Overall, looking at the impact of progressing trials and their response time, a significant effect is noticed. \textbf{(e)} The effect of error/correct answers with respect to the response time: Error answers result in longer response times.}
\end{figure}

\renewcommand{\thefigure}{S4}
\begin{figure}
     \centering
     \begin{subfigure}[b]{0.47\linewidth}
         \centering
         \includegraphics[width=\linewidth]{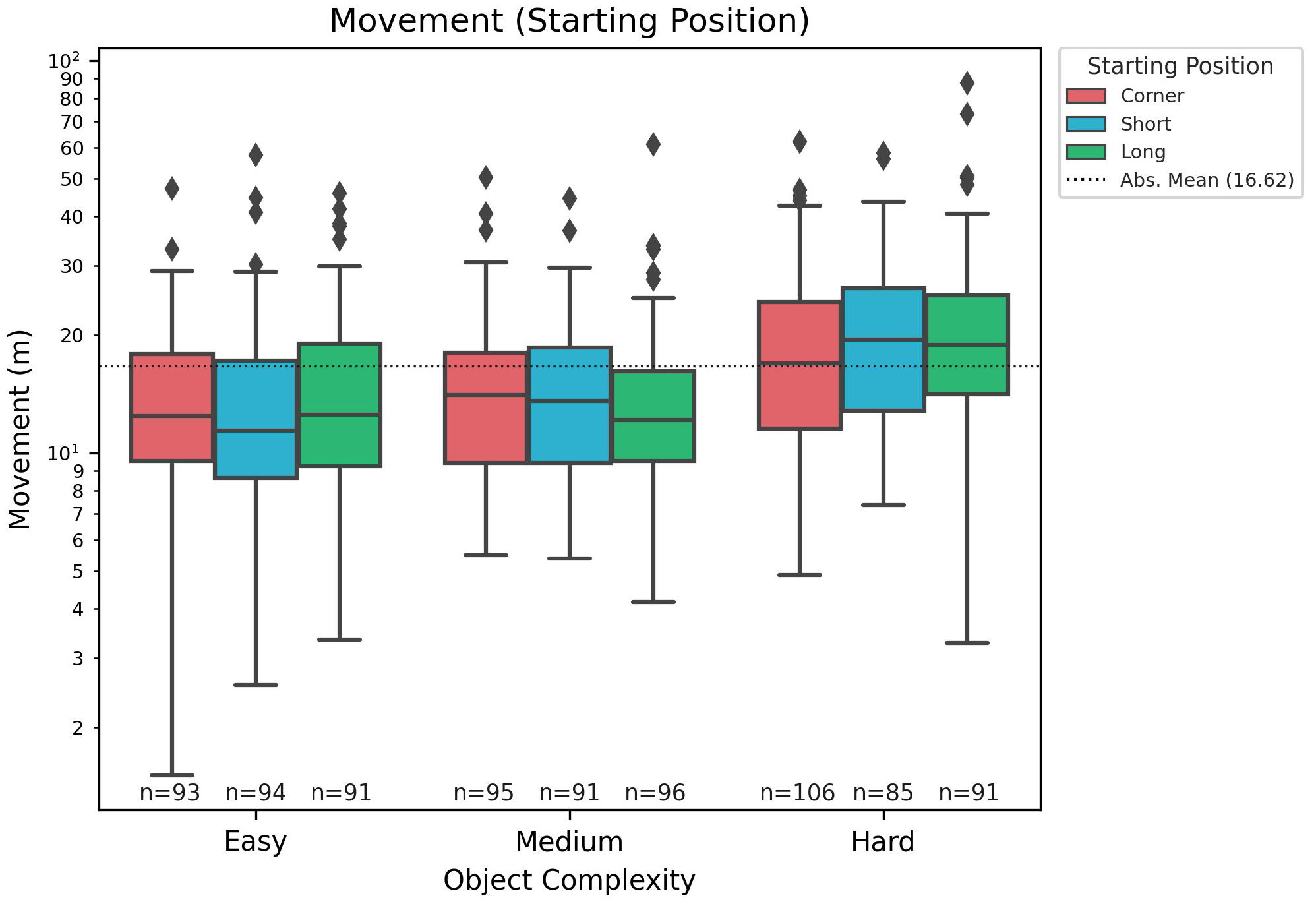}
         \caption{}
         \label{fig:res_move_pos}
     \end{subfigure}
     \hfill
     \begin{subfigure}[b]{0.47\linewidth}
         \centering
         \includegraphics[width=\linewidth]{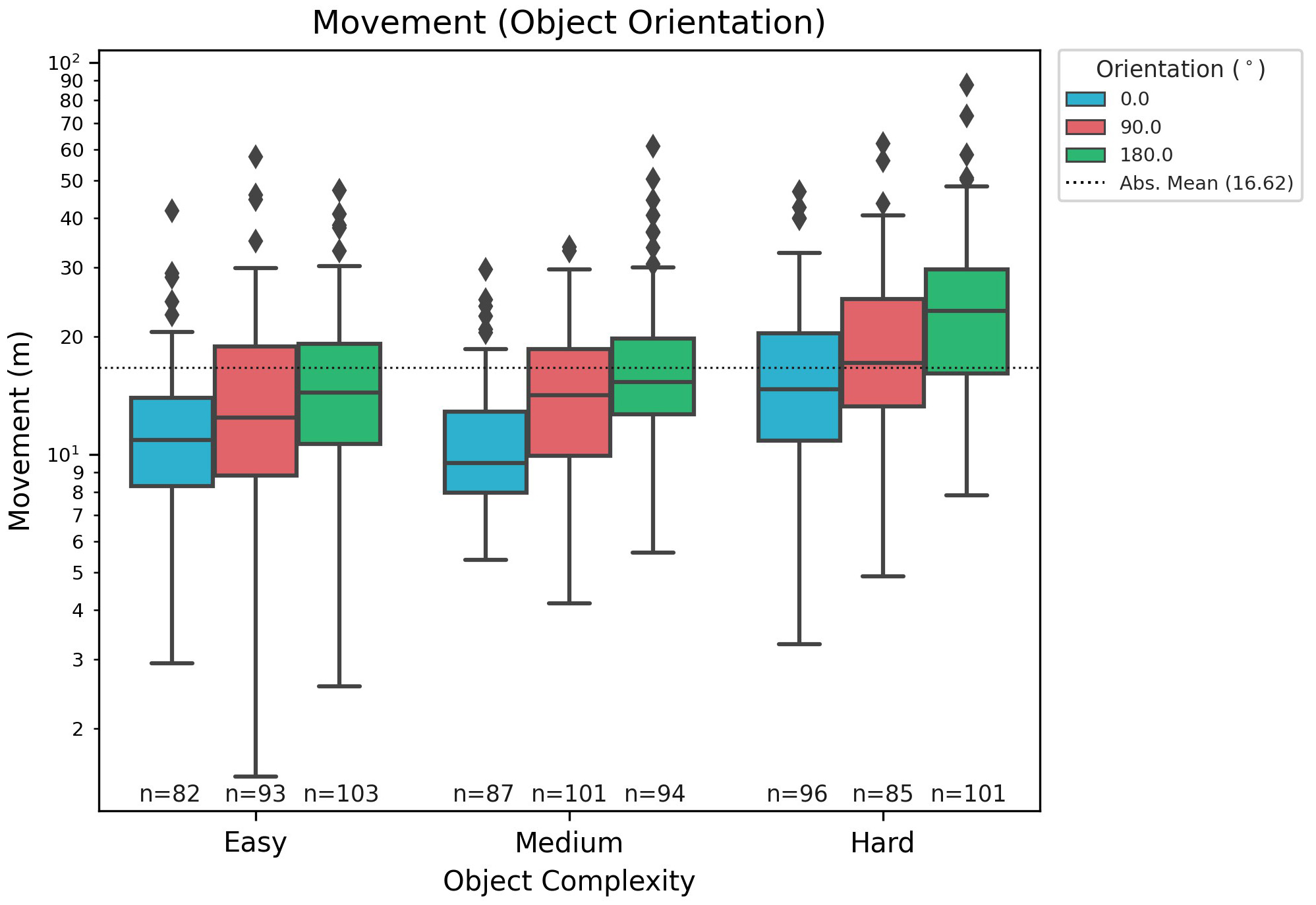}
         \caption{}
         \label{fig:res_move_orien}
     \end{subfigure}
     \hfill
     \begin{subfigure}[b]{0.47\linewidth}
         \centering
         \includegraphics[width=\linewidth]{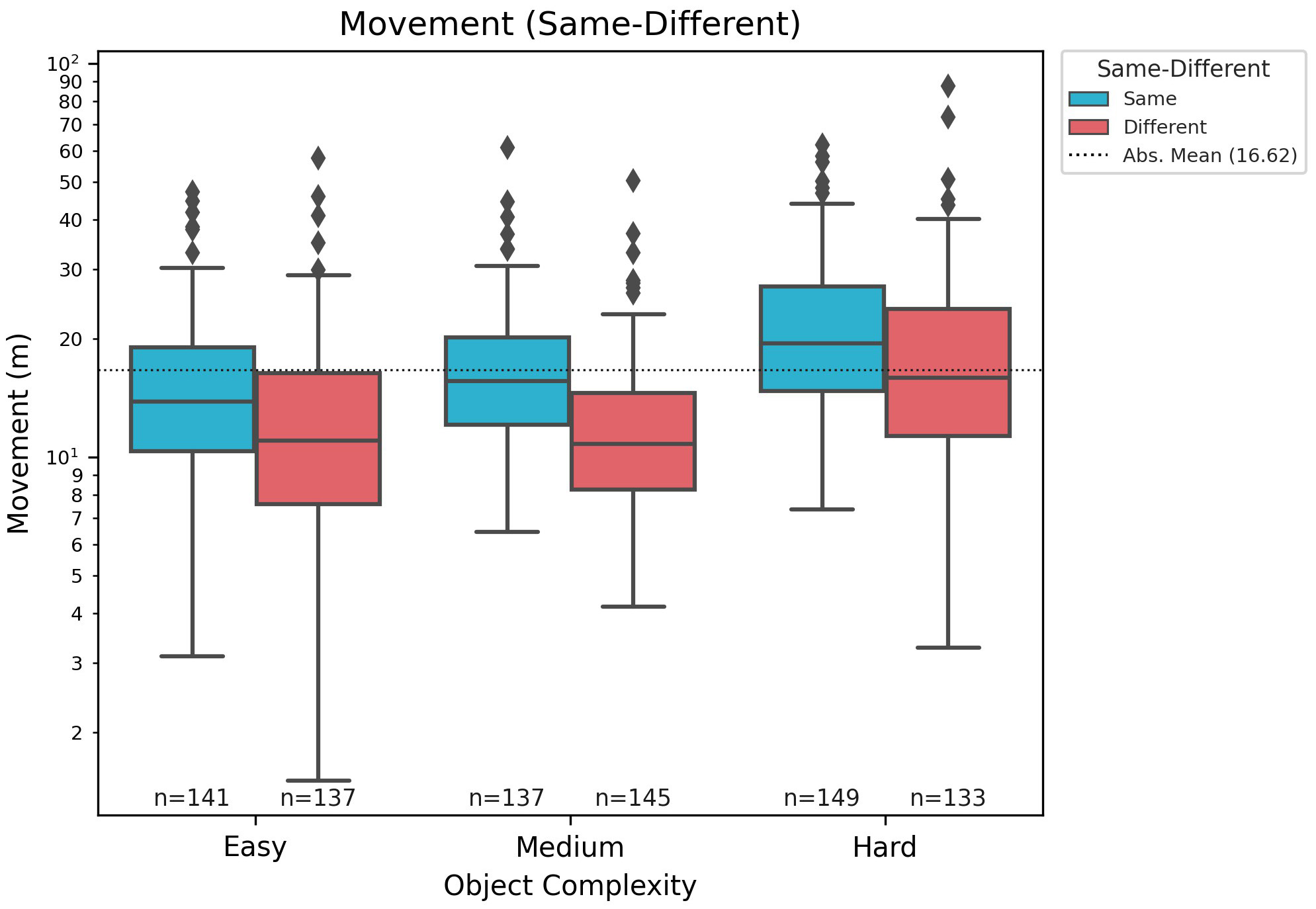}
         \caption{}
         \label{fig:res_move_same}
     \end{subfigure}
     \hfill
     \begin{subfigure}[b]{0.47\linewidth}
         \centering
         \includegraphics[width=\linewidth]{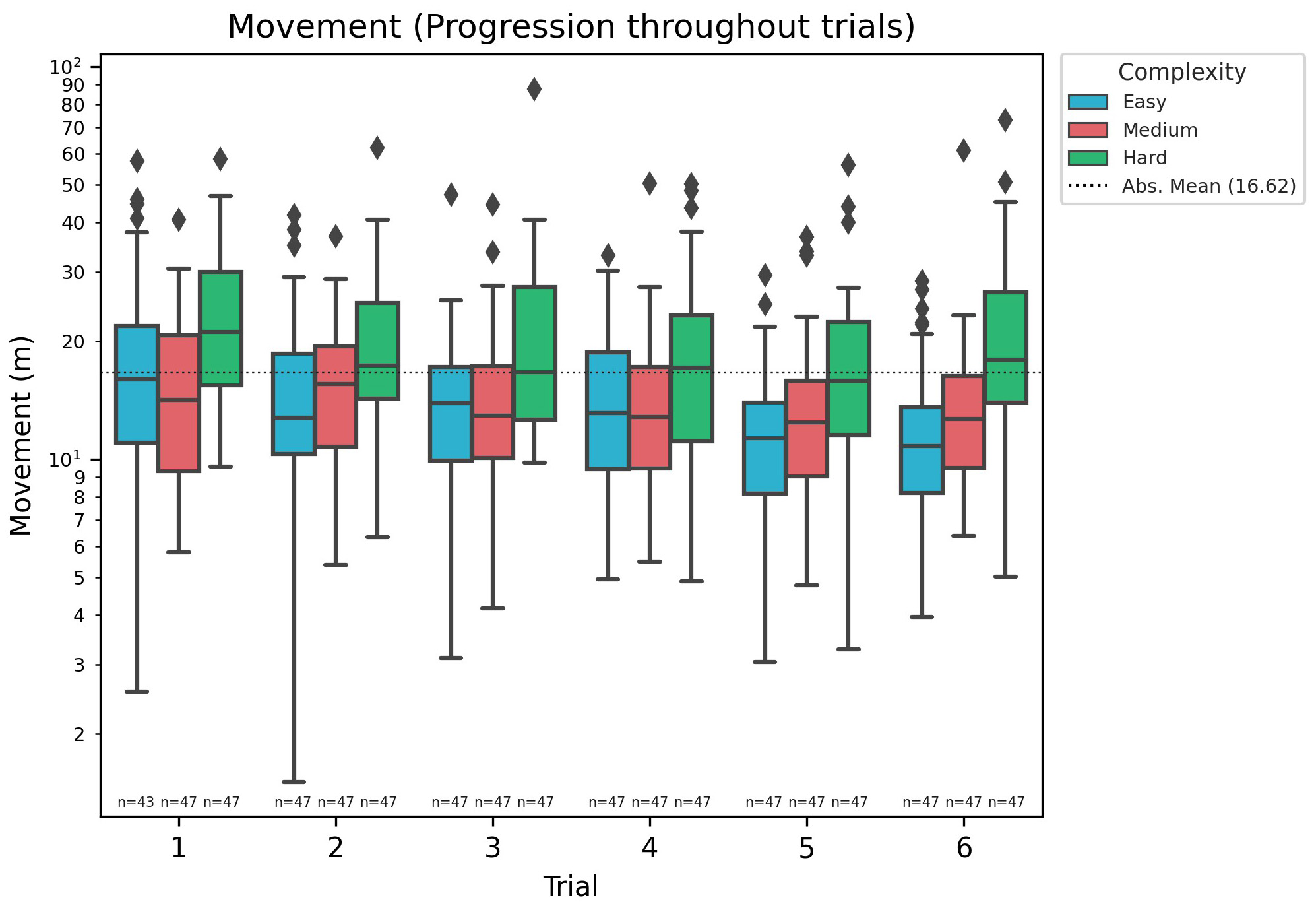}
         \caption{}
         \label{fig:res_move_progr}
     \end{subfigure}
     \hfill
     \begin{subfigure}[b]{0.47\linewidth}
         \centering
         \includegraphics[width=\linewidth]{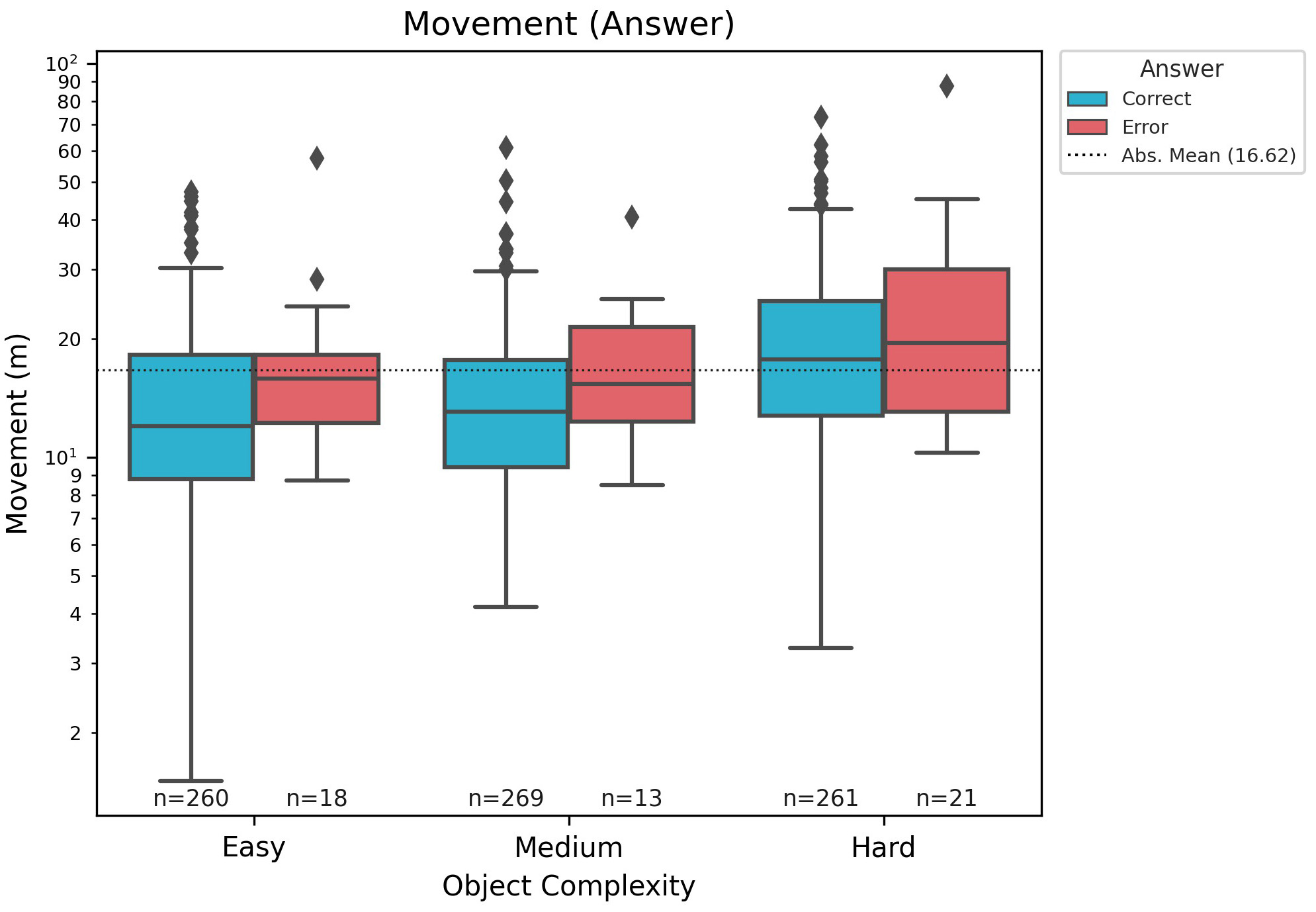}
         \caption{}
         \label{fig:res_move_answer}
     \end{subfigure}
     \caption{Head Movement against different experimental variables. \textbf{(a)} Starting position. We found no relationship between amount of head movement and starting position. \textbf{(b)} Object Orientation. A clear trend can be observed between the amount of head movement and the amount of orientational difference; at $0^{\circ}$ the least amount of movement was required, at $90^{\circ}$ an increase of 2-5m on average is recorded, and at $180^{\circ}$ an additional increase of 1-5m across all complexity classes is recorded. \textbf{(c)} Sameness. Aligned with the number of fixations, response time and accuracy, the amount of movement necessary is greater for the same object pairings across all complexity levels. For different cases, the increased upper and lower quartiles indicate that more uncertainty across different subjects in how to approach this case was involved. \textbf{(d)} Across trials. Visualizes the effect of the progression through trials. A significant reduction in head movement is noticable over the course of the trials. This means that participants show a learning effect in the sense that they execute a strategy with less head moment as the experiment progresses. For $C_m$ and $C_h$ cases, a trend is not visible directly, but for the $C_e$ cases, it is. $C_h$ cases start off at the first trial with just above 20m and drop to the absolute mean value of 16.62m and stay steady, marginally falling below and exceeding it repetitively; similarly, for the $C_m$ case, where no learning trend can be observed. However, the $C_e$ case, while noticing a slight up-trend for the second trial, consecutively decreases from about 16m down to about 10m, which is the equivalent of an improvement of 37.5\%. Lastly, \textbf{(e)} shows the effect of correct/error responses. However, there exists a significant effect on the correctness of the answer and head movement. Error responses were accompanied by significantly more head movement.}
     \label{fig:res_head_all}
\end{figure}

\renewcommand{\thefigure}{S5}
\begin{figure}
         \centering
         \includegraphics[width=1\linewidth]{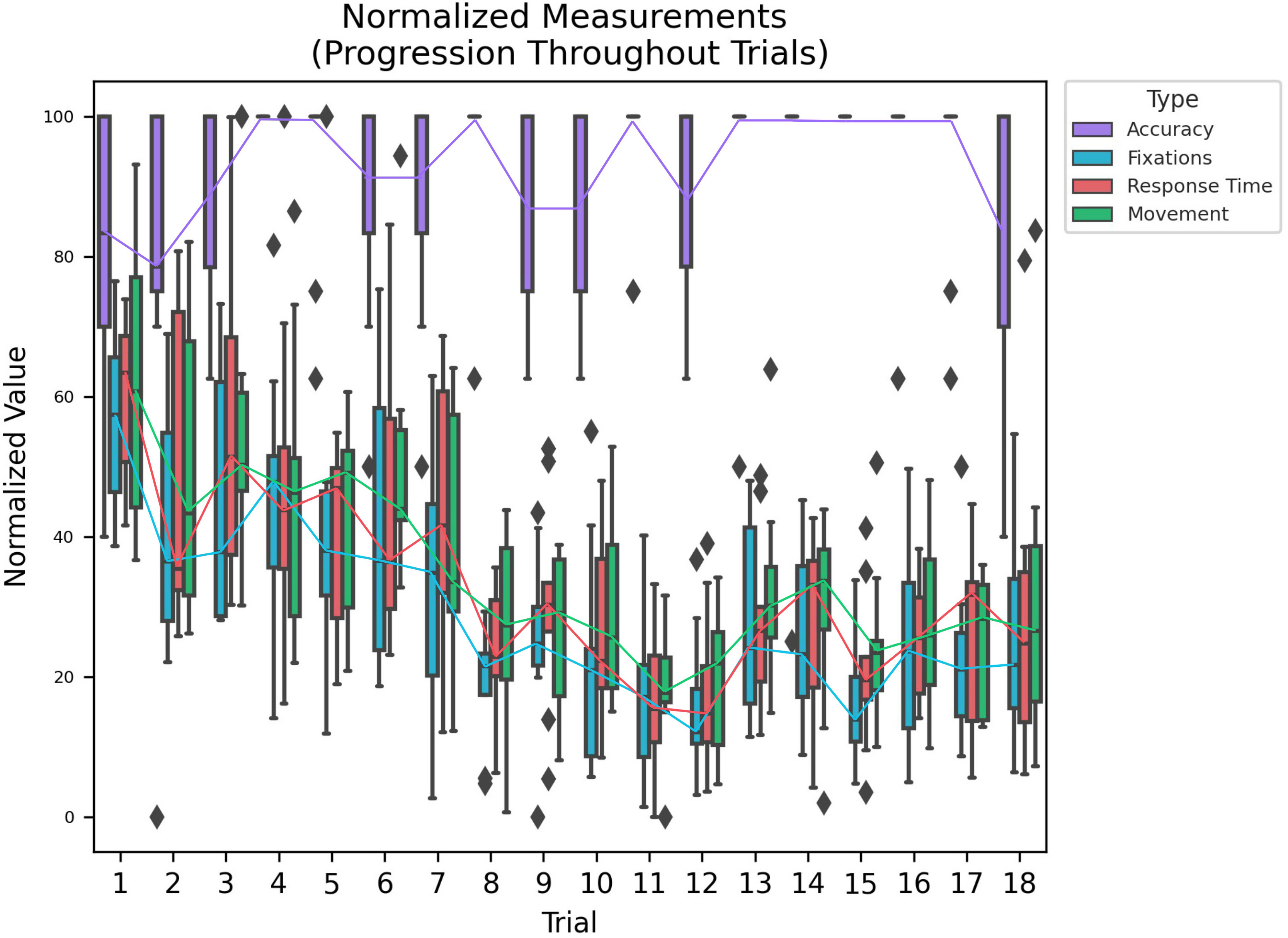}
         \caption{A plot of measured normalized variables (accuracy, fixations, response time and movement) with respect to trial number. It is easily seen that every variable improves over the course of the trials except for accuracy.}
         \label{fig:norm_all}
\end{figure}

\renewcommand{\thefigure}{S6}
\begin{figure}
     \centering
     \begin{subfigure}[b]{0.49\linewidth}
         \centering
         \includegraphics[width=\linewidth]{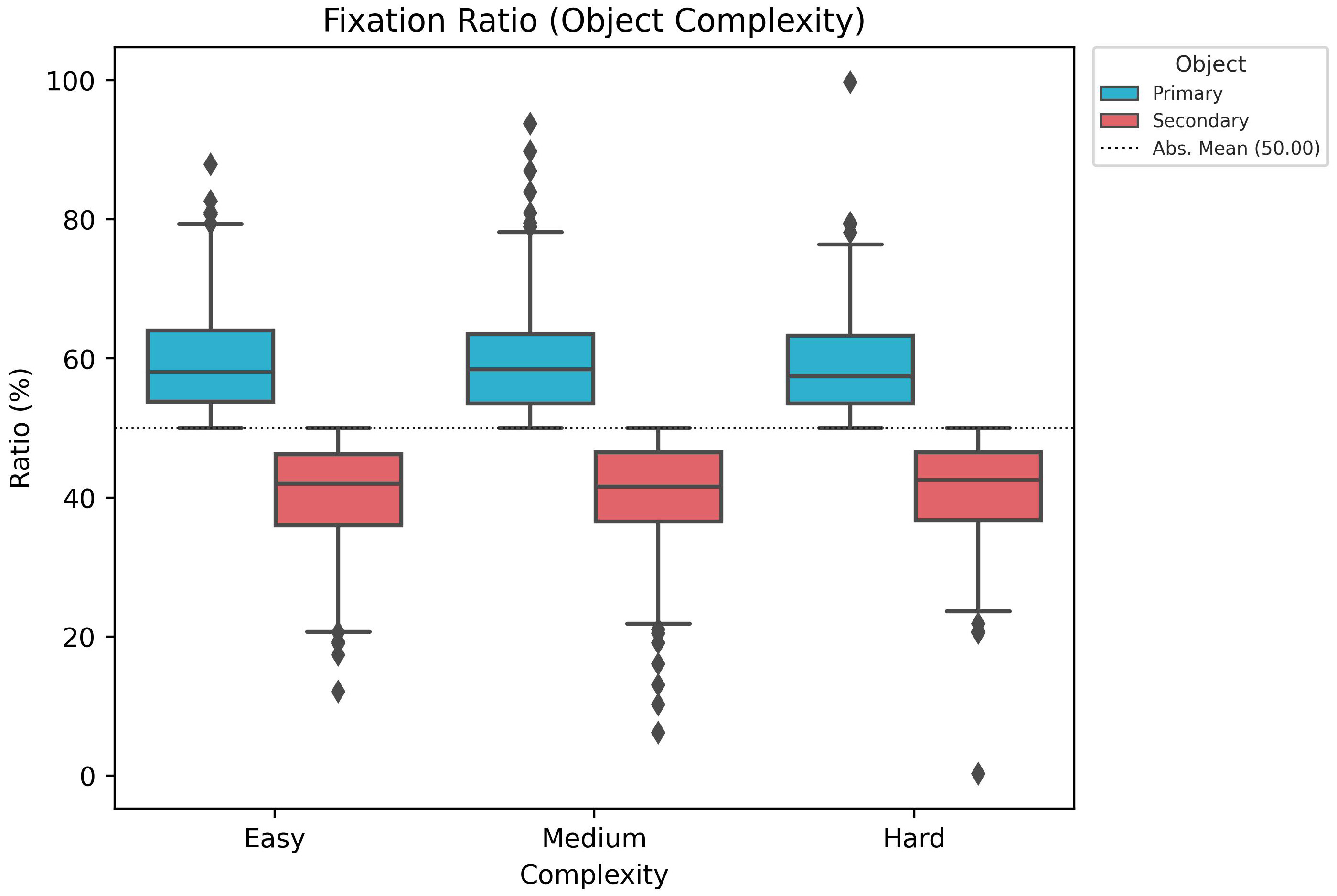}
         \caption{}
         \label{fig:res_move_orien}
     \end{subfigure}
     \hfill
     \begin{subfigure}[b]{0.39\linewidth}
         \centering
         \includegraphics[width=\linewidth]{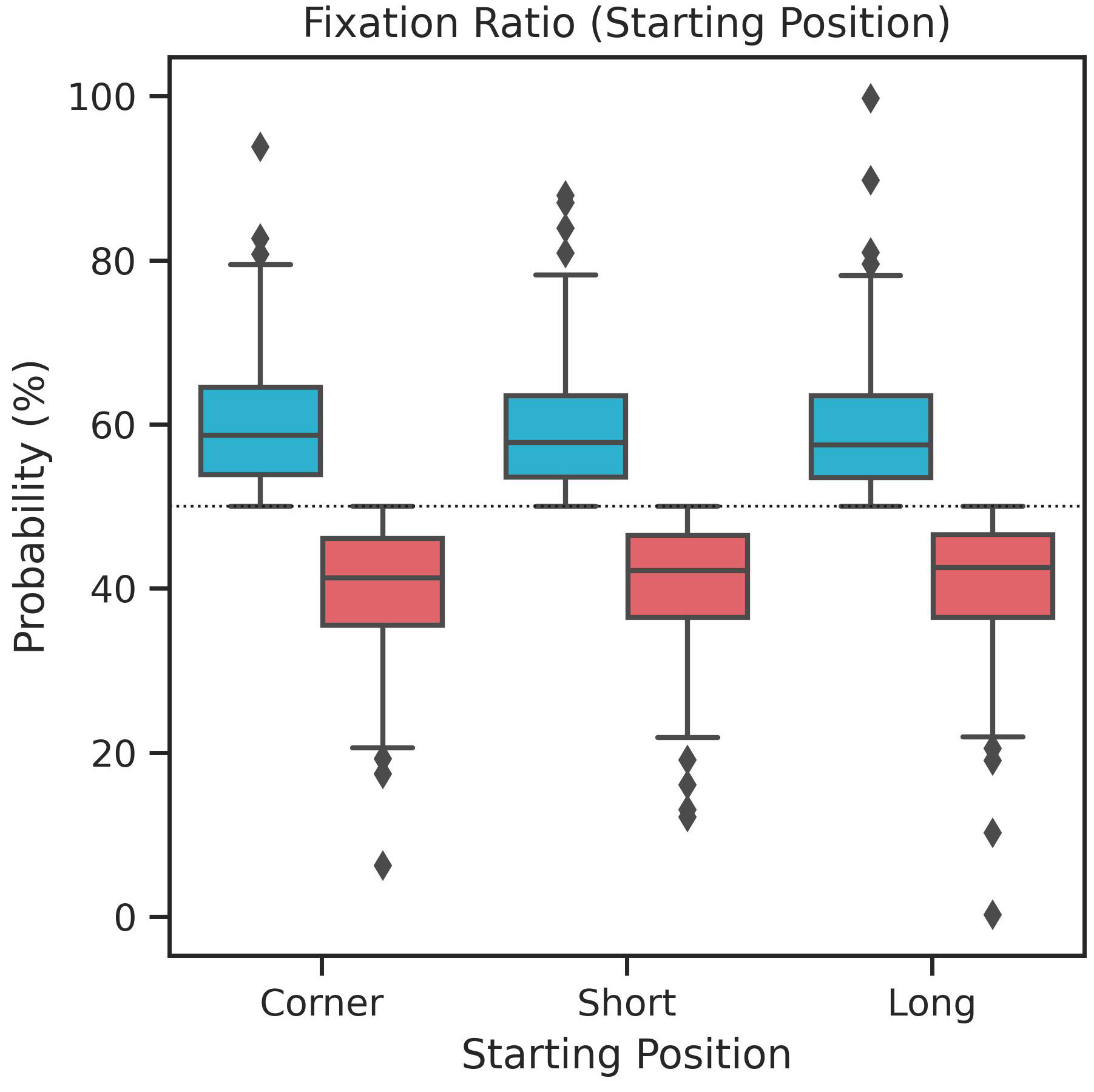}
         \caption{}
         \label{fig:res_move_orien}
     \end{subfigure}
     \hfill
     \begin{subfigure}[b]{0.39\linewidth}
         \centering
         \includegraphics[width=\linewidth]{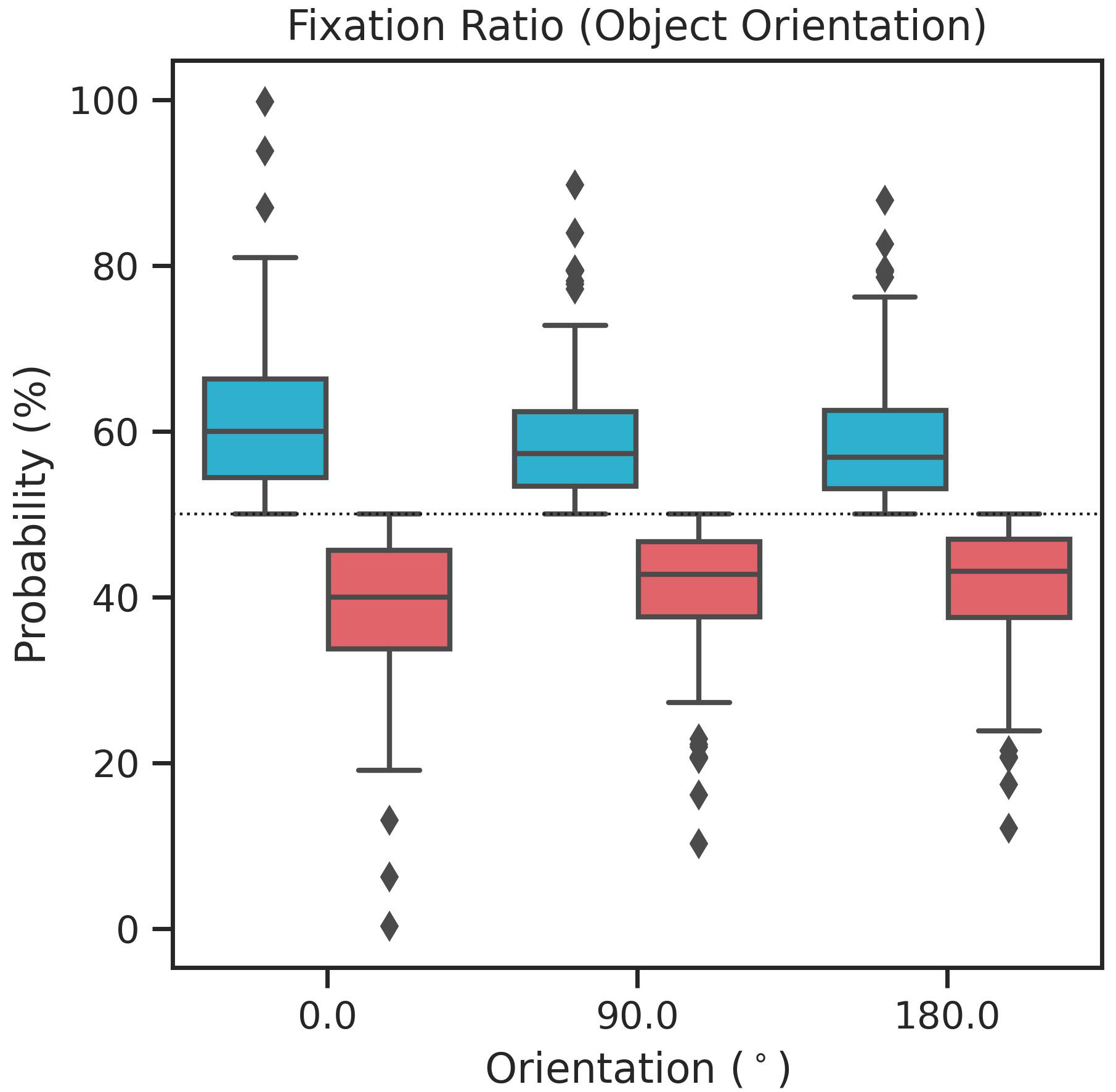}
         \caption{}
         \label{fig:res_move_same}
     \end{subfigure}
     \hfill
     \begin{subfigure}[b]{0.39\linewidth}
         \centering
         \includegraphics[width=\linewidth]{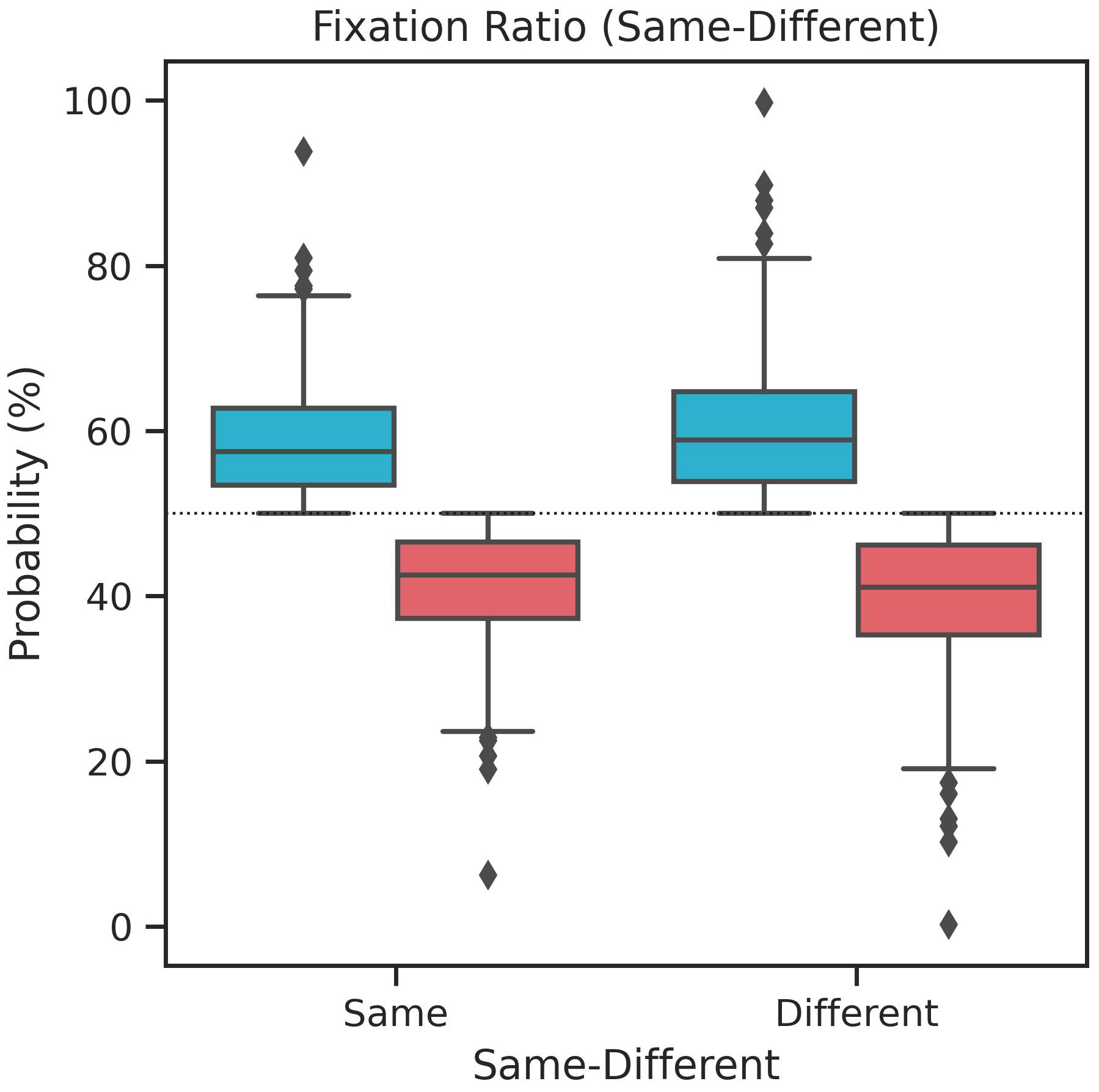}
         \caption{}
         \label{fig:res_move_progr}
     \end{subfigure}
     \hfill
     \begin{subfigure}[b]{0.39\linewidth}
         \centering
         \includegraphics[width=\linewidth]{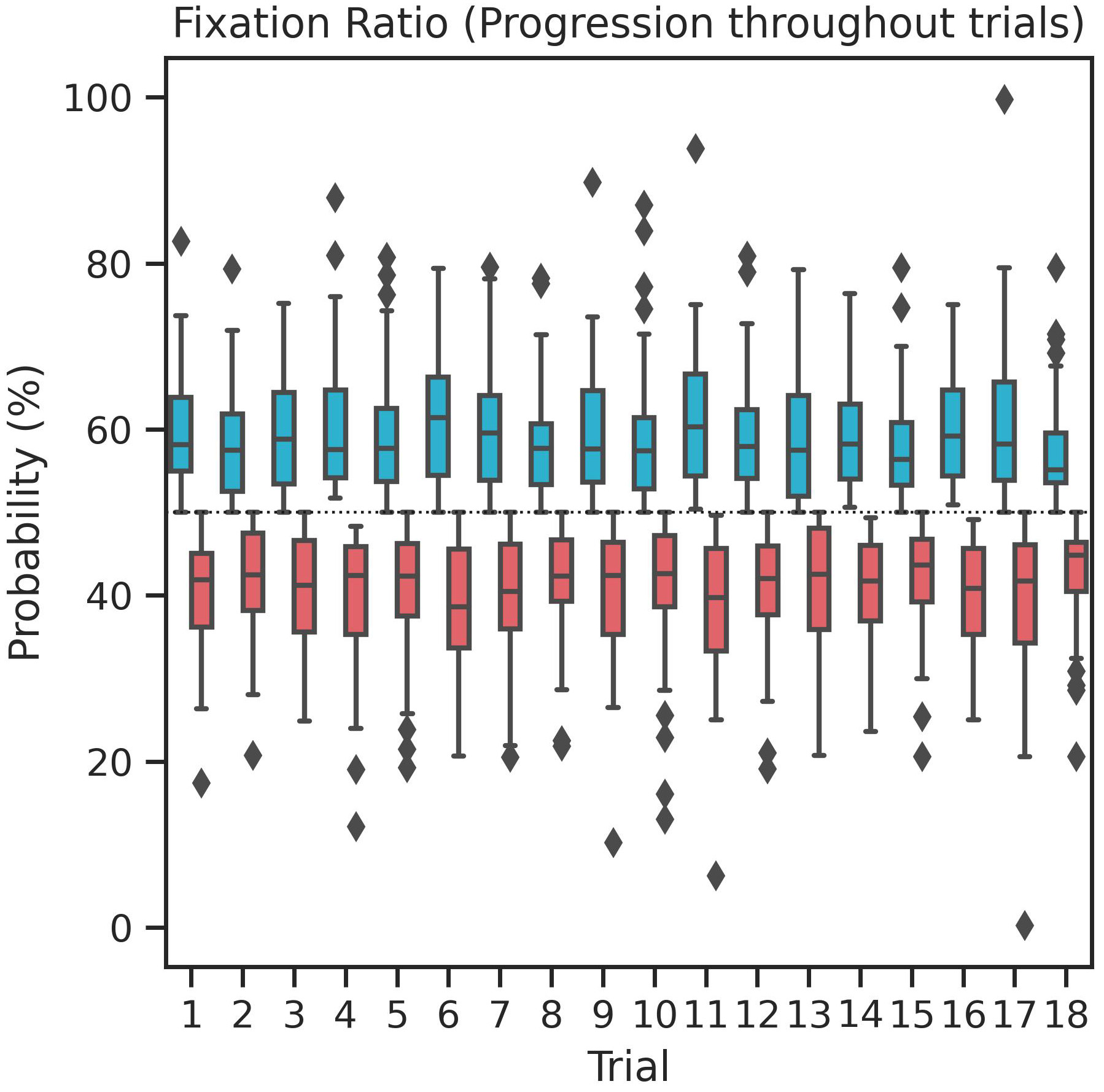}
         \caption{}
         \label{fig:res_move_answer}
     \end{subfigure}
     \hfill
     \begin{subfigure}[b]{0.39\linewidth}
         \centering
         \includegraphics[width=\linewidth]{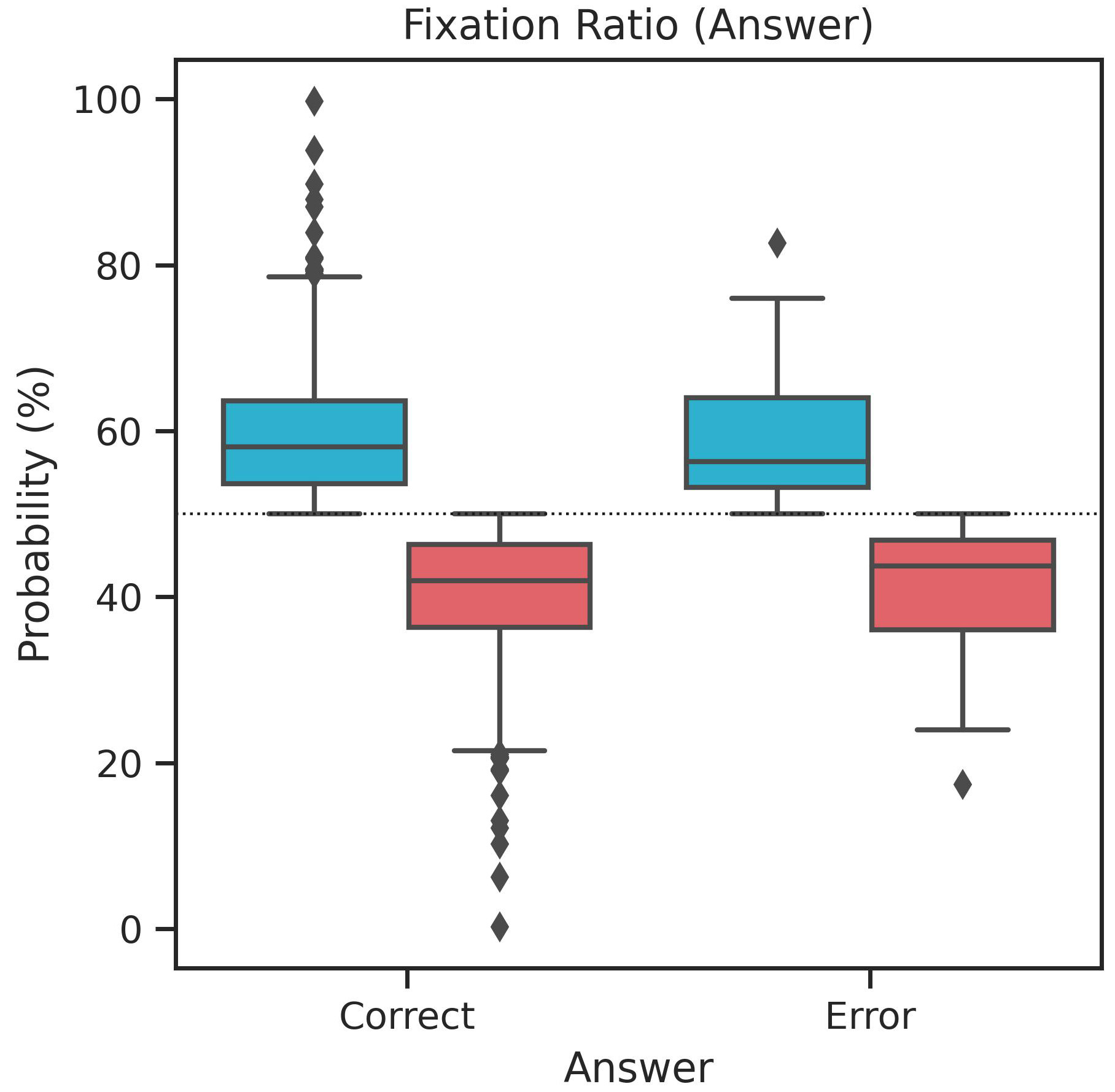}
         \caption{}
         \label{fig:res_move_answer}
     \end{subfigure}
     \caption{Ratio of fixations between both objects; \textbf{(a)} Object Complexity (this legend applies to all figures here), \textbf{(b)} Starting position, \textbf{(c)} Object Orientation, \textbf{(d)} Sameness, \textbf{(e)} Progression throughout trials, \textbf{(f)} Correct/Error Answer. We looked at the total number of fixations for either object. The object with the most fixations is considered the primary object, and the object with fewer or the same number of fixations is the secondary object. Interestingly, subjects tend to choose a primary and secondary object -- On average, the primary object accounted for $59.53\%$ of fixations, and the secondary object for $40.47\%$. However, none of the experimental variables have a significant effect on the fixation ratio.}
     \label{fig:res_head_all}
\end{figure}

\renewcommand{\thefigure}{S7}
\begin{figure}
     \centering
     \begin{subfigure}[b]{0.49\linewidth}
         \centering
         \includegraphics[width=\linewidth]{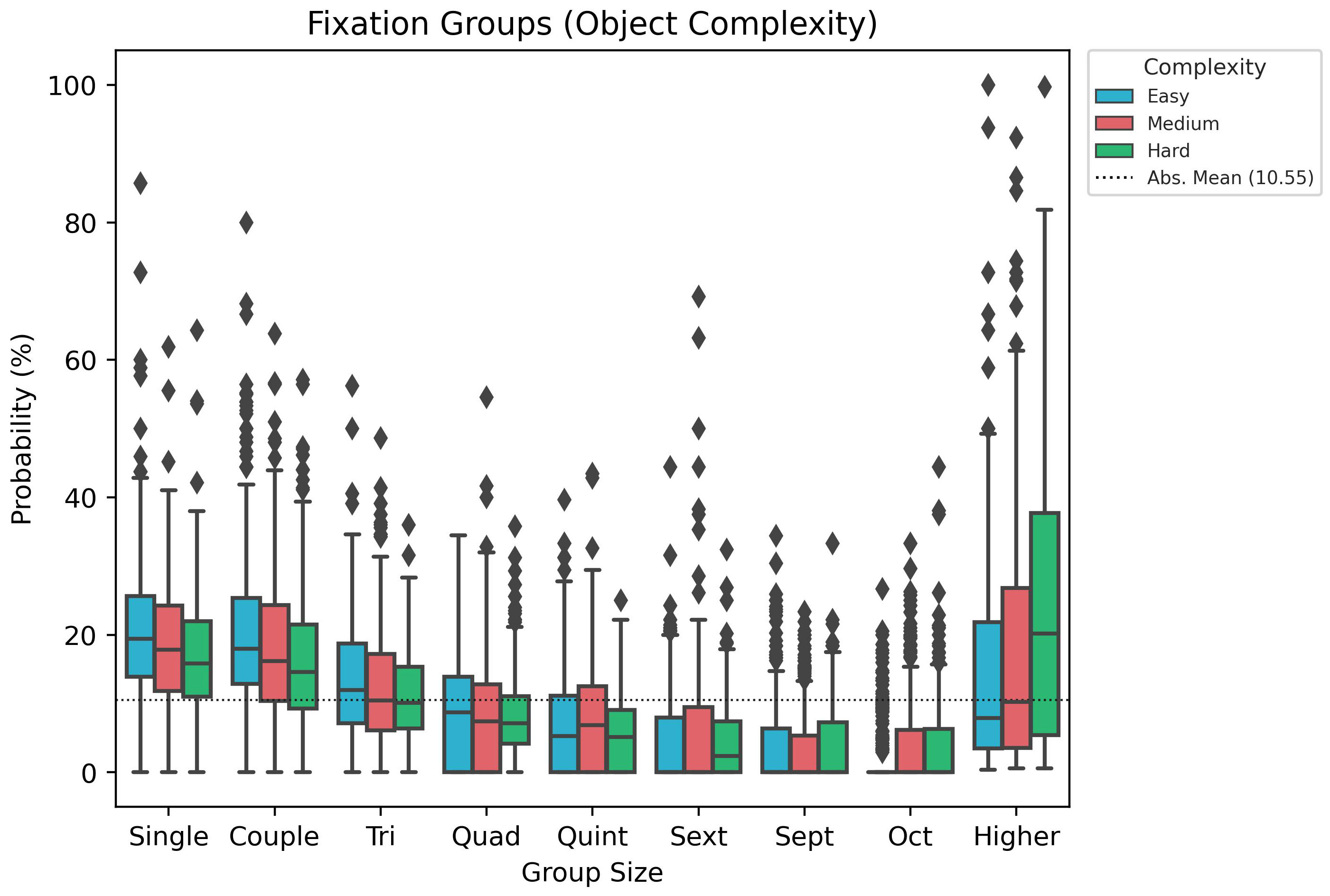}
         \caption{}
         \label{fig:res_move_orien}
     \end{subfigure}
     \hfill
     \begin{subfigure}[b]{0.49\linewidth}
         \centering
         \includegraphics[width=\linewidth]{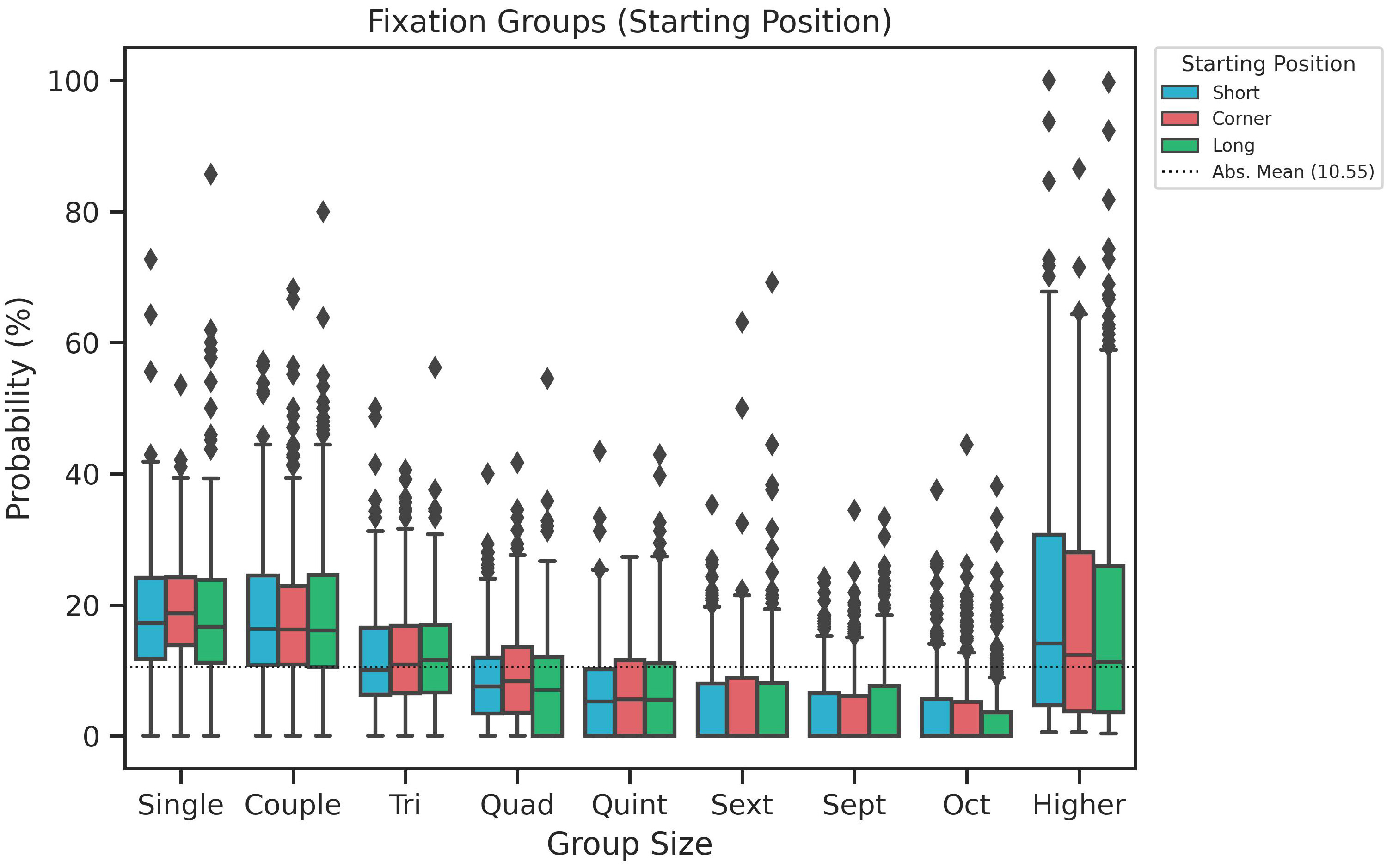}
         \caption{}
         \label{fig:res_move_orien}
     \end{subfigure}
     \hfill
     \begin{subfigure}[b]{0.49\linewidth}
         \centering
         \includegraphics[width=\linewidth]{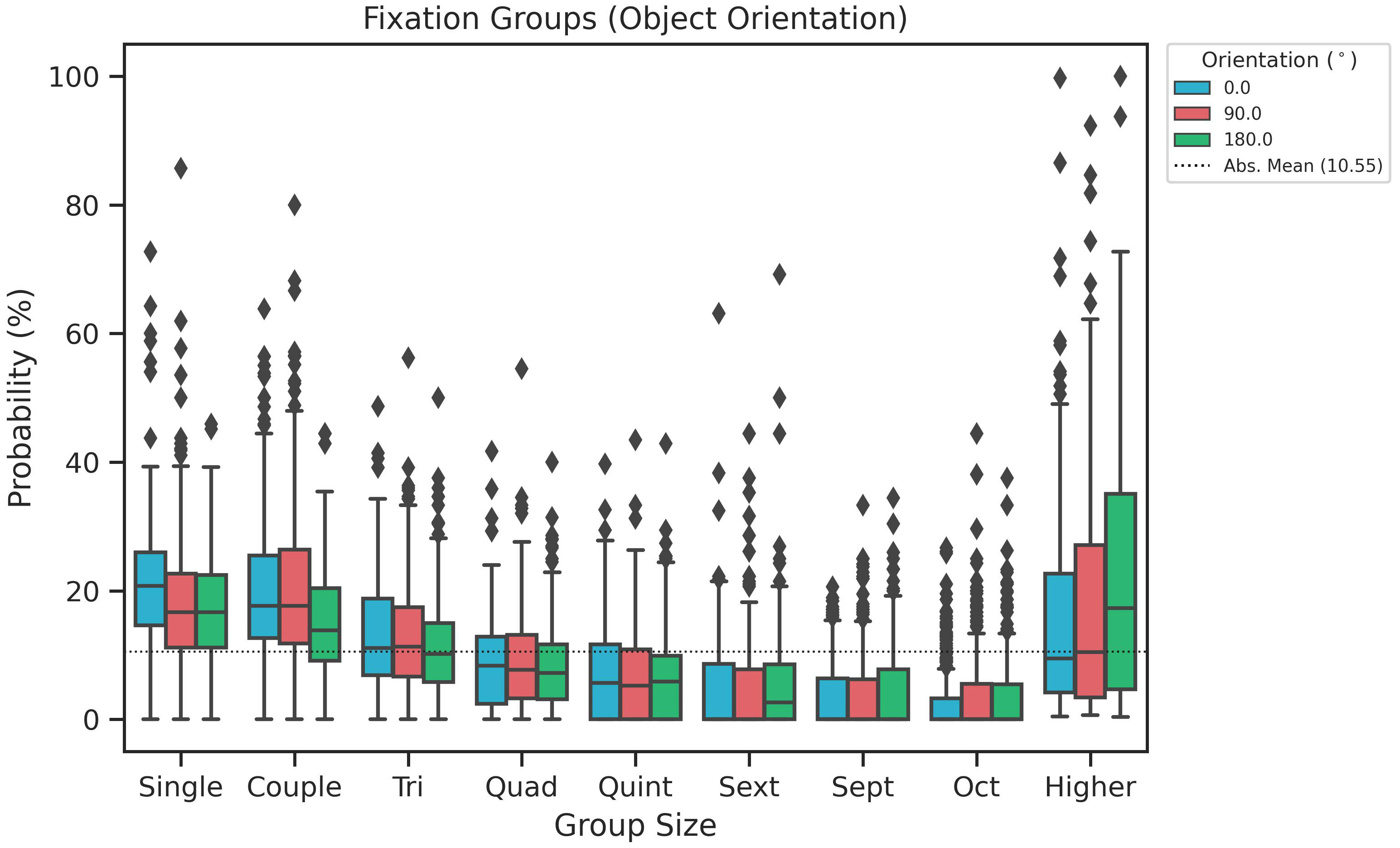}
         \caption{}
         \label{fig:res_move_same}
     \end{subfigure}
     \hfill
     \begin{subfigure}[b]{0.49\linewidth}
         \centering
         \includegraphics[width=\linewidth]{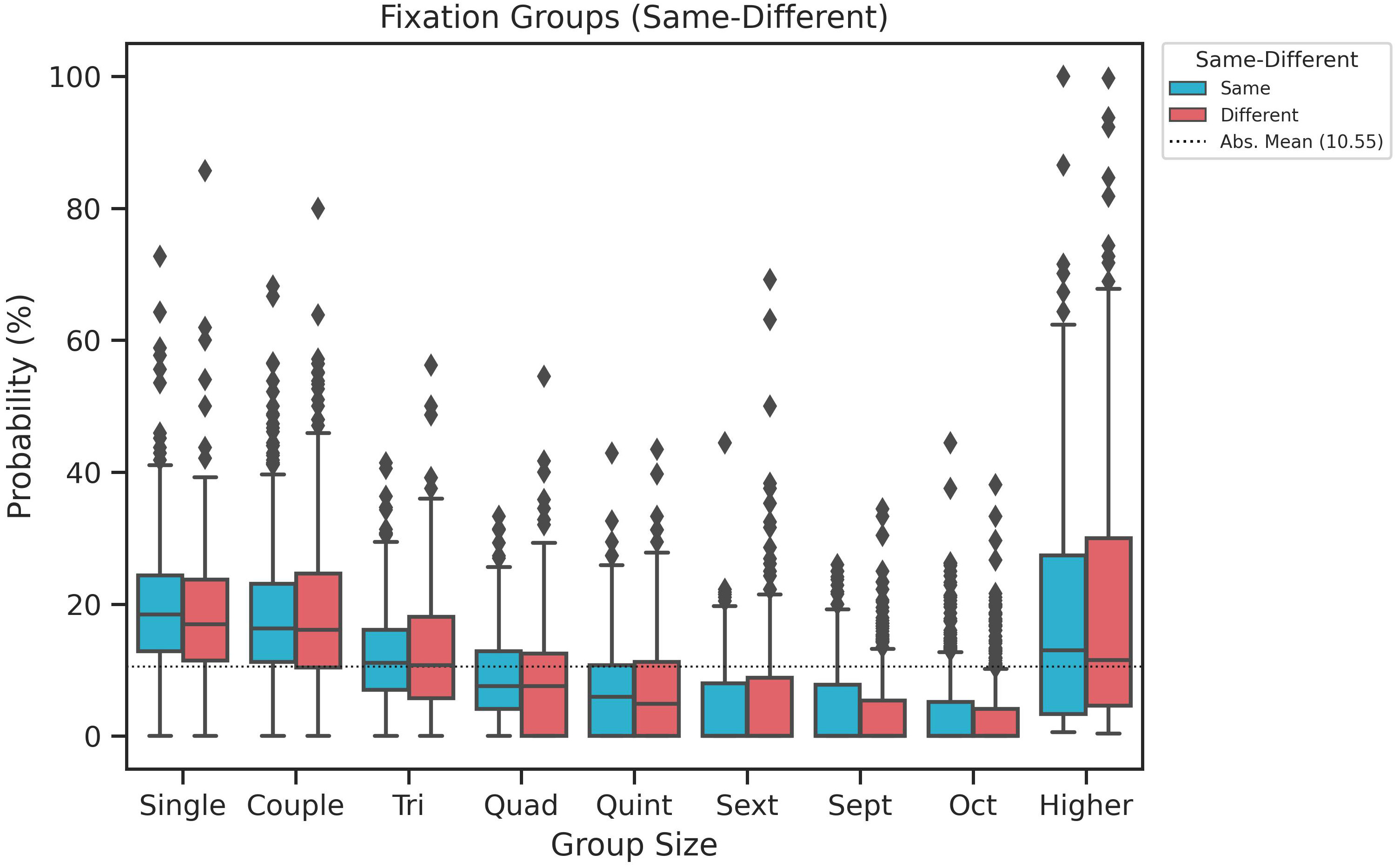}
         \caption{}
         \label{fig:res_move_progr}
     \end{subfigure}
     \hfill
     \begin{subfigure}[b]{0.49\linewidth}
         \centering
         \includegraphics[width=\linewidth]{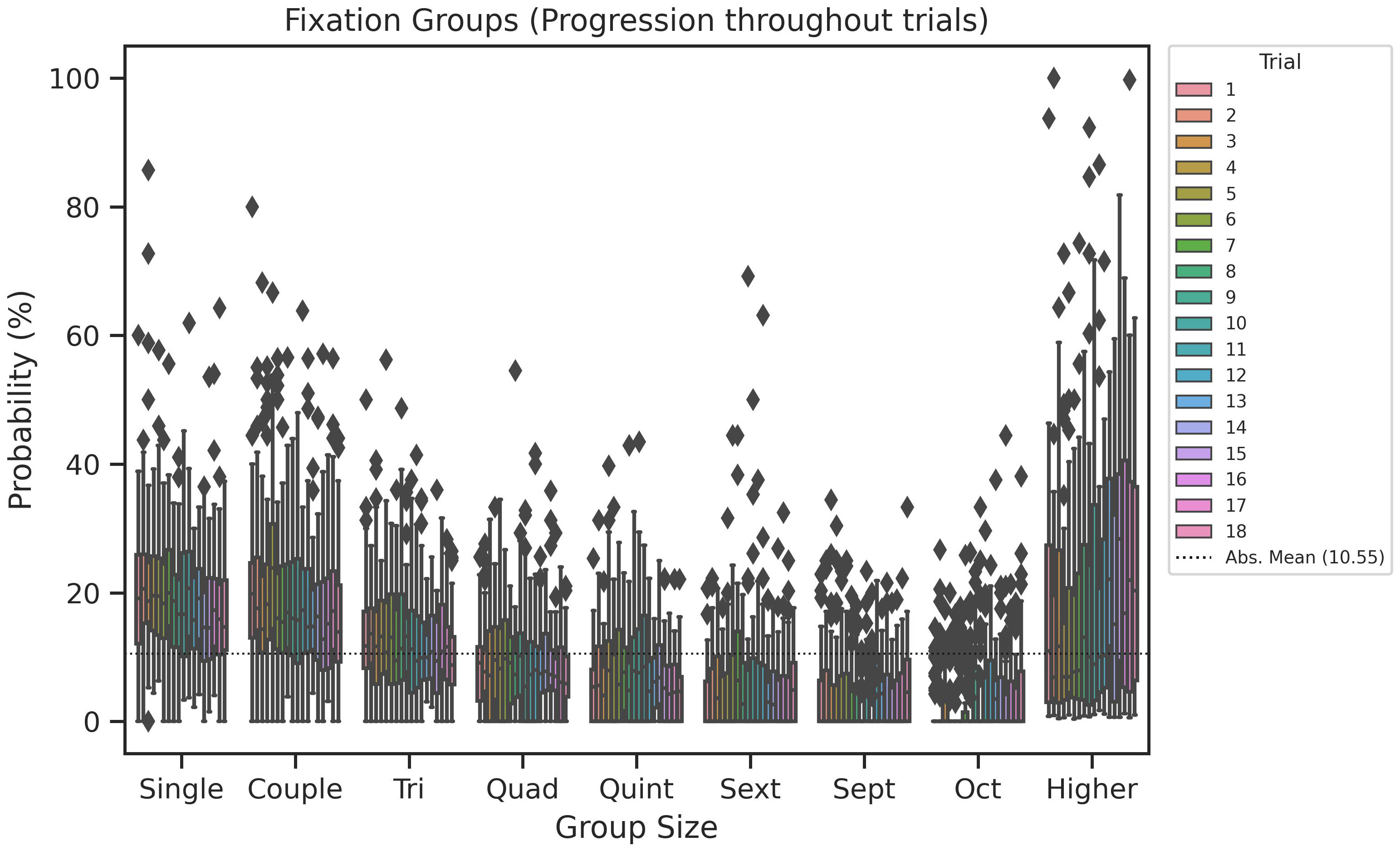}
         \caption{}
         \label{fig:res_move_answer}
     \end{subfigure}
     \hfill
     \begin{subfigure}[b]{0.49\linewidth}
         \centering
         \includegraphics[width=\linewidth]{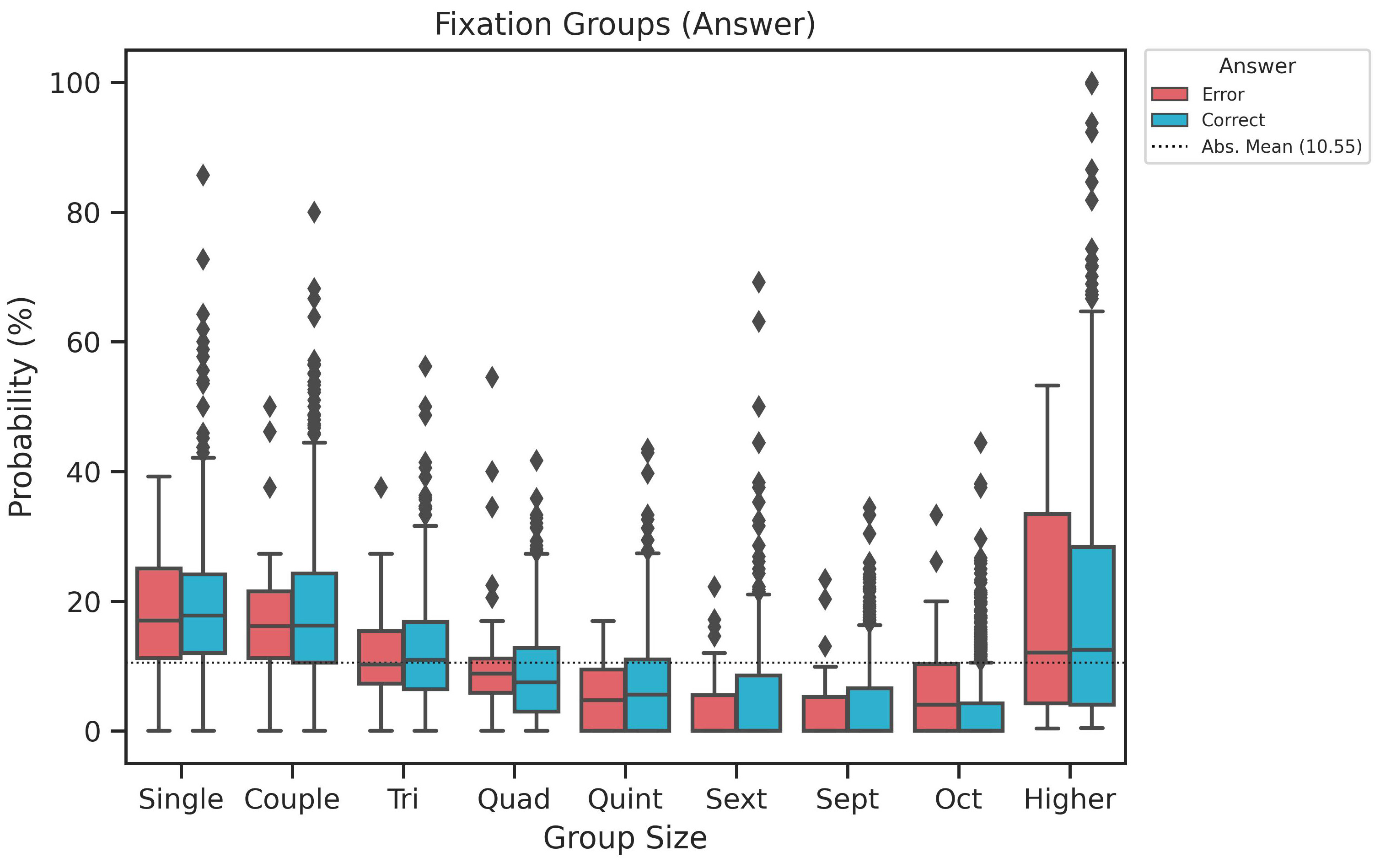}
         \caption{}
         \label{fig:res_move_answer}
     \end{subfigure}
     \caption{The result of the analysis of fixation groupings, i.e., the number of fixations on one object before changing focus to the other object. On average, $18.7\%$ ($\sigma = 10.01\%$) are fixations that change focus between each object every time. \textbf{(a)} The object complexity has an significant effect on single, couple, triple, quintuple, octuple and higher fixation groupings. Notably, for single, couple, and triple, the probability of occurrence significantly decreases with increasing object complexity. While for octuple and higher groupings, the opposite is true -- their occurrence increases with increasing object complexity. \textbf{(c)} The object orientation has a significant effect on single and couple groupings as this is the dominant method for object orientation $0^{\circ}$ and decreases steadily with increasing object orientation. Similar to object complexity, larger fixation groupings are affected by object orientation as well. Specifically, septuple and higher groupings occur more frequently with increasing object orientation. \textbf{(d)} The sameness of the object had largely no significant effect on fixation groups. Only single and septuple groups are more significantly used for the same objects than different ones. \textbf{(e)} As trials advanced, subjects used single groupings progressively less. A similar trend is observed for couple, triple, quadruple and quintuple groups -- none are significant, however. Larger groupings see an increase in probability as trials proceed. Notably, only a few sparse data points are recorded for octuple pairings up to trial 8. Octuple pairings are more regularly seen for trials 9-19. \textbf{(f)} The correctness of the answer significantly correlated with single and septuple pairs. Single and septuple fixation groups are significantly used more for correct answers than error responses.}
     \label{fig:ratio_fix_all}
\end{figure}

\renewcommand{\thefigure}{S8}
\begin{figure}
    \centering
    \includegraphics[width=1\linewidth]{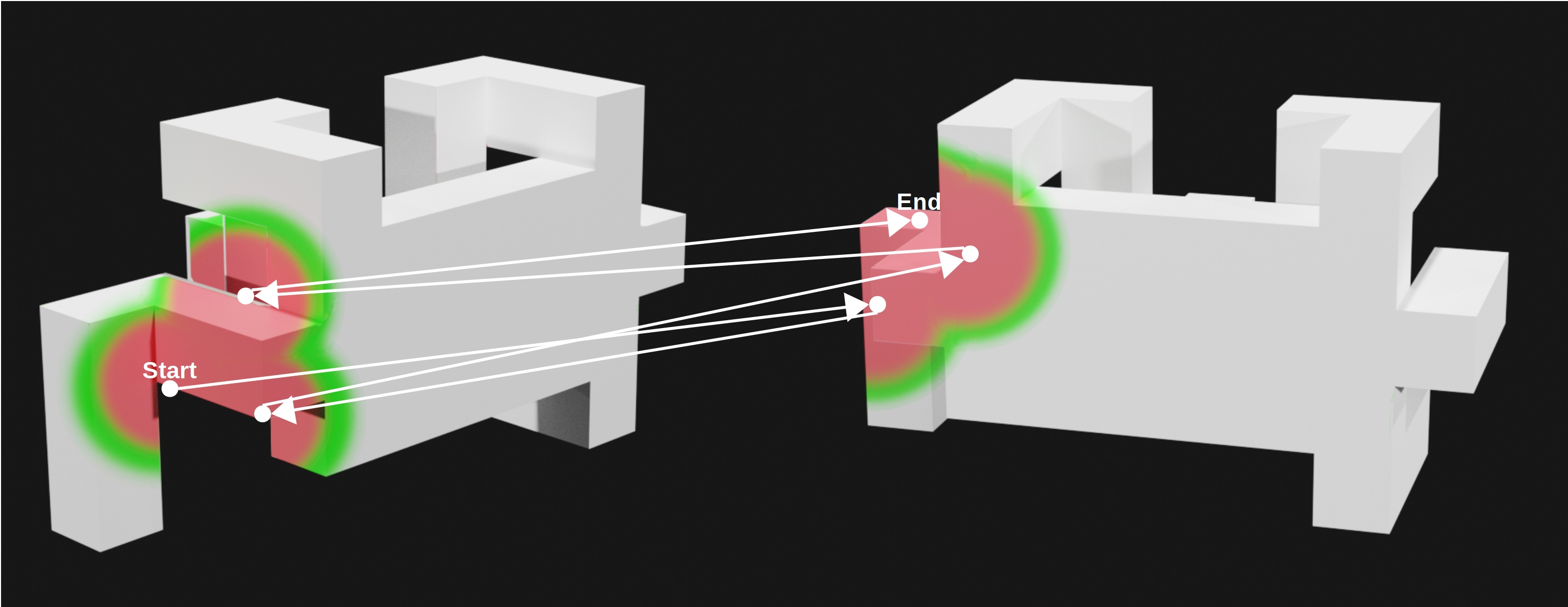}
    \caption{Here, we show a group of fixations that go back and forth between both stimulus. Both objects are displayed in the orientation of their observation. The corresponding fixations are highlighted with red circles and a green border. Arrows point to and originate at the center of gaze. Further, the starting fixation is provided (annotated with ``Start''), and the subsequent fixation is connected with an arrow. The alternating fixation ends at the fixation marked with ``End.'' The two objects are $C_m$; they are the same object, presented at $90^{\circ}$ orientational difference. The mean accuracy for gaze fixations is $1.42^{\circ}$. Color encoded with uncertainty boundary in green.}
    \label{fig:alter_fix}
\end{figure}

\end{document}